%% file: paper.tex
\documentclass[twocolumn]{aastex631}
\usepackage{amsmath}
\usepackage{CJKutf8}
\usepackage{makecell}
\usepackage{multirow}

\newcommand{\omc}{$\omega$\,Cen}
\newcommand{\ebv}{\ensuremath{E(B-V)}}
\newcommand{\nad}{\ensuremath{\mathrm{Na\,\textsc{i}\,D}}}
\newcommand{\nadtwo}{\ensuremath{\mathrm{Na\,\textsc{i}\,D_2}}}
\newcommand{\nadone}{\ensuremath{\mathrm{Na\,\textsc{i}\,D_1}}}
\newcommand{\snrnad}{$\rm{S/N_{\ensuremath{\mathrm{Na\,\textsc{i}\,D}}}}$}
\newcommand{\ewismdtwo}{$\rm{EW}_{\rm{D2, ISM}}$}
\newcommand{\ewismdone}{$\rm{EW}_{\rm{D1, ISM}}$}
\newcommand{\ewicmdtwo}{$\rm{EW}_{\rm{D2, ICM}}$}
\newcommand{\ewicmdone}{$\rm{EW}_{\rm{D1, ICM}}$}

\newcommand{\vlos}{$V_{\rm LOS}$}
\newcommand{\teff}{\ensuremath{T_{\mathrm{eff}}}}
\newcommand{\maspyr}{$\rm{mas}~\rm{yr}^{-1}$}
\newcommand{\perpixel}{$\text{pix}^{-1}$}

\newcommand{\angstrom}{\mbox{\normalfont\AA}}
\newcommand{\kms}{$\rm{km~s^{-1}}$}

\defcitealias{2023ApJ...958....8N}{N23}
\defcitealias{2024ApJ...970..192H}{H24}
\defcitealias{2010ApJ...710.1032A}{AvdM10}

\received{XX}
\revised{XX}
\accepted{XX}

\submitjournal{APJ}

\shorttitle{Interstellar and intracluster medium of \omc{}}
\shortauthors{Wang et al.}

\begin{document}
\begin{CJK}{UTF8}{gbsn}
\title{oMEGACat. VII. Tracing Interstellar and Intracluster Medium of $\omega$~Centauri using Sodium Absorptions}

\correspondingauthor{Zixian Wang (Purmortal, 王梓先)}
\email{wang.zixian.astro@gmail.com}

\author[0000-0003-2512-6892]{Z. Wang (王梓先)}
\affiliation{Department of Physics and Astronomy, University of Utah, Salt Lake City, UT 84112, USA}

\author[0000-0003-0248-5470]{A. C. Seth}
\affiliation{Department of Physics and Astronomy, University of Utah, Salt Lake City, UT 84112, USA}

\author[0000-0002-7547-6180]{M. Latour}
\affiliation{Institut für Astrophysik und Geophysik, Georg-August-Universität Göttingen, Friedrich-Hund-Platz 1, 37077 Göttingen, Germany}

\author[0000-0002-1468-9668]{J. Strader}
\affiliation{Center for Data Intensive and Time Domain Astronomy, Department of Physics and Astronomy, Michigan State University, 567 Wilson Road, East Lansing, MI 48824, USA}

\author[0000-0002-5844-4443]{M. H\"aberle}
\affiliation{Max Planck Institute for Astronomy, K\"onigstuhl 17, D-69117 Heidelberg, Germany}

\author[0000-0002-6922-2598]{N. Neumayer}
\affiliation{Max Planck Institute for Astronomy, K\"onigstuhl 17, D-69117 Heidelberg, Germany}


\author[0009-0005-8057-0031]{C. Clontz}
\affiliation{Department of Physics and Astronomy, University of Utah, Salt Lake City, UT 84112, USA}
\affiliation{Max Planck Institute for Astronomy, K\"onigstuhl 17, D-69117 Heidelberg, Germany}

\author[0000-0001-6604-0505]{S. Kamann}
\affiliation{Astrophysics Research Institute, Liverpool John Moores University, 146 Brownlow Hill, Liverpool L3 5RF, UK}

\author[0000-0002-2941-4480]{M. S. Nitschai}
\affiliation{Max Planck Institute for Astronomy, K\"onigstuhl 17, D-69117 Heidelberg, Germany}

\author[0000-0002-1212-2844]{M. Alfaro-Cuello}
\affiliation{Centro de Investigación en Ciencias del Espacio y Física Teórica, Universidad Central de Chile, La Serena 1710164, Chile}

\author[0000-0003-3858-637X]{A. Bellini}
\affiliation{Space Telescope Science Institute, 3700 San Martin Dr., Baltimore, MD, 21218, USA}

\author[0000-0002-0160-7221]{A. Feldmeier-Krause}
\affiliation{Department of Astrophysics, University of Vienna, T\"urkenschanzstrasse 17, 1180 Wien, Austria}

\author[0000-0001-9673-7397]{M. Libralato}
\affiliation{INAF - Osservatorio Astronomico di Padova, Vicolo dell'Osservatorio 5, Padova I-35122, Italy}

\author[0000-0001-7506-930X]{A. P. Milone}
\affiliation{Dipartimento di Fisica e Astronomia “Galileo Galilei”, Universita’ di Padova, Vicolo dell’Osservatorio 3, Padova, IT-35122}
\affiliation{Istituto Nazionale di Astrofisica - Osservatorio Astronomico di Padova, Vicolo dell’Osservatorio 5, Padova, IT-35122}

\author[0000-0002-7489-5244]{P. J. Smith}
\affiliation{Max Planck Institute for Astronomy, K\"onigstuhl 17, D-69117 Heidelberg, Germany}
\affiliation{Fakultät für Physik und Astronomie, Universität Heidelberg, Im Neuenheimer Feld 226, D-69120 Heidelberg, Germany}

\author[0000-0001-8052-969X]{S. O. Souza}
\affiliation{Max Planck Institute for Astronomy, K\"onigstuhl 17, D-69117 Heidelberg, Germany}

\author[0000-0003-4546-7731]{G. van de Ven}
\affiliation{Department of Astrophysics, University of Vienna, T\"urkenschanzstrasse 17, 1180 Wien, Austria}



\begin{abstract}

We investigate the foreground interstellar medium along the line of sight and intracluster medium of $\omega$~Centauri (\omc{}) by measuring the equivalent width of \nad{} absorptions from MUSE observations.
The large line-of-sight velocity difference between \omc{} and the foreground enables us to separate \nad{} absorption contributed from atomic gas in the interstellar and intracluster medium.
We find that small-scale substructures in the foreground \nad{} distribution correlate with differential reddening derived from photometric methods.
Using an empirical \nad{} equivalent width-reddening relation, we determine an average reddening of $E(B-V)=0.153\pm0.003$~mag within the half-light radius of \omc{}.
However, the \nad{}-inferred differential reddening is significantly larger than photometric estimates. This is likely due to scatter in the \nad{}-reddening relation.
We find no evidence for intracluster atomic gas from spectra of horizontal branch stars, as there is no significant \nad{} absorption at \omc{}'s systemic velocity.
Given this non-detection, we place the strongest upper limit to date on the intracluster atomic gas column density in \omc{} of $\lesssim2.17 \times 10^{18}~\rm{cm^{-2}}$.
We also estimate the ionized gas density from pulsar dispersion measure variations, which exceed the atomic gas limit by $\sim$50 times.
Nevertheless, the strong correlation between dispersion measure and foreground \nad{} suggests that much or all of this ionized gas resides in the foreground.
Given ongoing mass loss from bright giant stars, our findings imply that the intracluster gas accumulation timescale is short, and gas removal in the cluster is likely not tied to stripping as \omc{} passes through the Galactic disk.

\end{abstract}



\section{Introduction} 
\label{sec:intro}

For over 50 years, the content of the intracluster medium (ICM) of globular clusters has been an open question. 
Early on, the focus was on searching for the accumulated gas from ``extra" mass loss that could reconcile main sequence turn-off masses with the lower inferred masses of horizontal branch stars \citep{Iben1971}. 
Such searches placed tight limits on the presence of neutral \ion{H}{1} in clusters \citep{Knapp1973}. 
While the existence of this post-main-sequence integrated mass loss is now well-established (e.g., \citealt{Howell2022}), it is uncertain what stellar phases dominate the mass loss, and if this differs for low-mass metal-poor stars compared to stars of higher mass and metallicity that are better studied.

The presence of ICM in globular clusters also has broader implications beyond stellar mass loss.
For example, if intermediate-mass black holes are present in some clusters, then they should accrete the ambient ICM and become visible as low-luminosity active galactic nuclei. 
Therefore, if typical globular clusters contain ICM, then radio or X-ray studies can set limits on the presence of intermediate-mass black holes \citep{Grindlay2001, Maccarone2004, Strader2012, 2013ApJ...773L..31H, 2018ApJ...862...16T}. 

There are at least two challenges in searching for the ICM in globular clusters. First, it can exist in multiple phases: (i) ionized gas, detectable via pulsar dispersion measures and $H_\alpha$ emission; (ii) atomic gas\footnote{In this study, use the term ``atomic gas'' to refer to non-ionized gas that is often traced by neutral hydrogen measurements.}, revealed through optical spectral features and radio HI line emission;  (iii) dust, which affects photometric measurements, and (iv) molecular gas, potentially detectable via mm/radio spectroscopy. 
The second challenge is that an ICM signal can be hidden by the stronger signal of the foreground multi-phase interstellar medium (ISM).

The most widely observed component of the ISM is dust.
Its average reddening and extinction can be constrained through isochrone fitting, while its spatial variation manifests as differential reddening. 
Dust can influence the analysis of multiple stellar populations, as well as the determination of ages and metallicities \citep[e.g.,][]{2012A&ARv..20...50G, 2013MNRAS.435..263B, 2015AJ....149...91P, 2017MNRAS.464.3636M}. 
The most common approach to mapping differential reddening in globular clusters involves measuring stars' displacement along the reddening vector, relative to a defined fiducial sequence on the color-magnitude diagram (CMD).
This approach assumes that such displacements are primarily caused by reddening variations \citep[e.g.,][]{1973AJ.....78..597T, 1999AJ....118.1727P, 2001AJ....121.1522V, 2011AJ....141..146A}, although instrumental effects in photometry can also contribute.
This technique has been extensively applied to photometric data from both HST and ground-based telescopes \citep[e.g.,][]{2013MNRAS.435..263B, 2022MNRAS.517.5687J, 2023MNRAS.522..367L, 2024AA...686A.283P, 2024A&A...690A..37S}. 
Recent studies have shown that differential reddening correction effectively reduces the color broadening of CMD sequences, enabling more robust identification of elusive features - such as multiple stellar populations in globular clusters, blue stragglers and binary systems \citep[e.g.,][]{2012ApJ...751L...8P, 2017ApJ...844..164B}. 
However, a major challenge in using dust to study the ICM is separating any potential signal from the cluster itself from the (typically stronger) foreground interstellar medium (ISM) contribution, which is often difficult and relies on many assumptions.

An alternative approach is based on spectroscopy. 
Several studies identified spectral features associated with interstellar gas that can be used to probe the different compositions of the ISM \citep[e.g.,][]{2001ApJ...552L..73A, 2009MNRAS.399..195V, 2015MmSAI..86..527M, 2017A&A...607A.133W}. 
Among these, the \nad{} absorption that originates from atomic sodium gas is one of the most widely used tracers of the neutral ISM \citep{1974ApJ...191..381H, 1994AJ....107.1022R, 1997A&A...318..269M}. 
An empirical relation between the equivalent width (EW) of \nad{} and the color excess \ebv{} has been established for low-resolution spectra by \cite{2012MNRAS.426.1465P} and \cite{2015MNRAS.452..511M}.
This empirical relation provides a useful method to estimate reddening directly from spectroscopic observations.
An advantage in using optical spectral lines is that they can be matched to the cluster velocity to separate ICM from ISM \citep[e.g.,][]{2009MNRAS.399..195V}.

Over the past two decades, many studies have explored the ISM and ICM of \omc{}, the most massive star cluster in the Milky Way with a distance of $5.43 \pm 0.05$~kpc \citep{2021MNRAS.505.5957B}. 
These studies focused on differential reddening \citep{2005ApJ...634L..69C, 2017ApJ...842....7B, 2024ApJ...970..192H}, diffuse interstellar bands (DIBs), and the presence of cold and warm gas traced by \nad{} and $\mathrm{Ca\,\textsc{ii}\,K}$ absorption \citep{2009MNRAS.399..195V}. 
By applying differential reddening corrections, \cite{2017ApJ...844..164B} successfully identified 15 subpopulations in \omc{} using HST photometry.

However, several questions remain unanswered regarding the intracluster and foreground gas content of \omc{}.
First, \cite{2024ApJ...977...14C} measured the ages of sub-giant branch (SGB) stars within $\sim5'$ of \omc{} and found that a mean value of $E(B-V) = 0.185$~mag was required to match the main-sequence turn-off (MSTO) regions on the CMD to other low extinction clusters with similar metallicity.
This value is higher than previously reported estimates \citep[e.g.,][]{2009MNRAS.394..831M, 2010arXiv1012.3224H, 2023A&A...677A..86L}; it is unclear whether this variation is due to field-of-view differences or differences in methodology.  
Second, no clear detection of intracluster gas has yet been made in \omc{}. 
This is particularly interesting because of the recent identification of fast-moving stars at the cluster center, which implies the presence of an intermediate-mass black hole with a mass between 8200 and 50,000~M$_\odot$ \citep[][]{2024Natur.631..285H}. 
However, this black hole was not detected in previous deep X-ray and radio measurements \citep{2013ApJ...773L..31H,2018ApJ...862...16T}.  
\citet{2018ApJ...862...16T} presented modeling of the radio measurements assuming an ionized ICM similar to that measured from pulsars in 47~Tuc \citep{2001ApJ...557L.105F}, and found that any black hole above 1000~M$_\odot$ should have been detectable.  
Therefore, the lack of such detection may imply a lack of gas at the cluster center, which is unusual given the expected mass loss from post-main-sequence stars as discussed above. Clearly, better constraints on the internal gas density of \omc{} are needed. 

Several studies have attempted to investigate the ICM and mass loss of \omc{} using \textit{Spitzer} photometry \citep[e.g.,][]{2008AJ....135.1395B, 2009MNRAS.394..831M}, but the relative contributions of different intracluster gas components and whether the results were contaminated by foreground material remain unclear.
\cite{2009MNRAS.399..195V} separated \nad{} absorptions originating from intracluster and foreground cold sodium gas of \omc{} by using the line-of-sight velocity (\vlos{}) difference of these two components. 
They found no significant evidence of ICM from their averaged spectra, but did not provide quantitative upper limits on the ICM component.
Their study focused on 452 blue-HB stars spanning a field of view with a radius of $0.5^{\circ}$, making it difficult to resolve small-scale structures in the cluster’s inner region.
As a result, the composition and distribution of intracluster gas within the half-light radius of \omc{} ($\sim5'$) remain poorly constrained.
With the newly compiled oMEGACat photometric and spectroscopic datasets from HST and MUSE \citep{2023ApJ...958....8N, 2024ApJ...970..192H}, which fully cover \omc’s half-light radius, we now have the opportunity to extend the work of \cite{2009MNRAS.399..195V} and investigate the ISM and ICM of \omc{} using \nad{} absorptions in greater detail.

\begin{figure}
    \centering
    \includegraphics[width=1\linewidth]{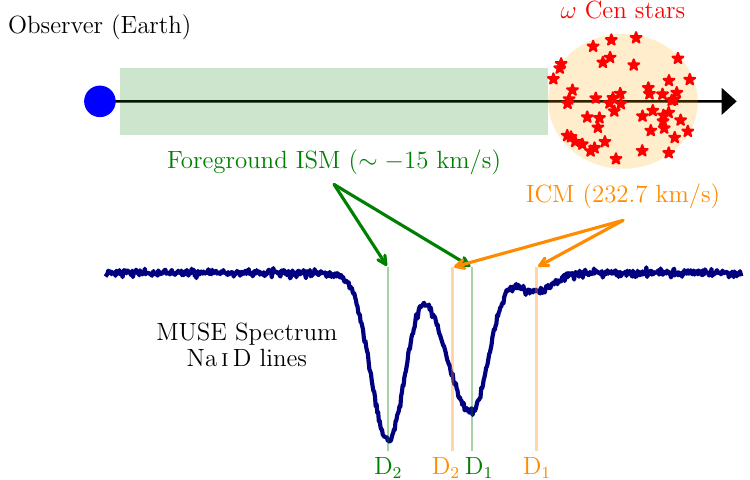}
    \caption{
    An illustration of the method used in this paper to probe the foreground interstellar medium (ISM) and intracluster medium (ICM) using the \nad{} region of our MUSE spectra.  The top portion shows the locations of the observer, the intervening interstellar medium, and the intracluster medium; each is labeled with its radial velocity.  These differing radial velocities result in a Doppler offset between the interstellar and intracluster \nad{} lines as shown in the bottom panel, which shows where we expect to see the lines from each component in our spectra.  
    }
    \label{fig:scratch_ism_icm}
\end{figure}

In this paper, we follow the approach of \cite{2009MNRAS.399..195V}, using \nad{} absorption features from MUSE observations to trace the ISM and ICM separately.
As illustrated in Fig.~\ref{fig:scratch_ism_icm}, this is achieved due to the significant line-of-sight velocity difference between \omc{} (the systemic velocity of 232.7~\kms{} is used throughout this work, \citealt{2018MNRAS.478.1520B}) and foreground stars ($-5$ to $-25$~\kms{}), which creates a \nad{} wavelength difference of $\lambda>4.58$~\AA{}.
This difference allows us to distinguish \nad{} absorption originating from intracluster and foreground sodium atomic gas under MUSE spectral resolution (see the bottom panels of Fig.~\ref{fig:samplefig}).
Using our measurements, we infer the foreground gas distribution and reddening distribution within the half-light radius and constrain the intracluster atomic gas density.
This sample also enables the study of mass loss using bright giant stars.

We describe the oMEGACat datasets and the data preprocessing in Section~\ref{sec:data}.
In Section~\ref{sec:methods}, we describe our algorithm for measuring the equivalent width of \nad{} absorption and converting it to reddening.
In Section~\ref{sec:results-ism}, we present the spatial distribution of \nad{} equivalent width from foreground gas, and then explore the \nad{} absorption due to intracluster gas in Section~\ref{sec:results-icm}.
In Section~\ref{sec:disscuss}, we compare our inferred small-scale interstellar gas structures and differential reddening with previous photometric studies, constrain the intracluster gas column density and its timescale of stellar mass loss, and investigate potential connections between foreground ISM and \omc{}.
Finally, we provide a summary in Section~\ref{sec:summary}.


\section{Data} 
\label{sec:data}

\subsection{oMEGACat with Other Datasets}
\label{subsec:data-muse-hst}

We use datasets from the oMEGACat project\footnote{\url{https://omegacatalog.github.io/}}, which aims to disentangle the dynamics and formation history of \omc{} by compiling the largest spectroscopic, photometric, and astrometric dataset to date.
The data consists of two main catalogs covering \omc{} within the half-light radius ($\sim5'$): a spectroscopic catalog derived from over 100 MUSE pointings on \omc{}, containing more than 300,000 individual stellar spectra with metallicity and line-of-sight velocity (\vlos{}) measurements (\citealt{2023ApJ...958....8N}, hereafter \citetalias{2023ApJ...958....8N}); and a HST-based catalog containing proper motions and photometry in 7 bands for up to $\sim1.4$ million stars (\citealt{2024ApJ...970..192H}, hereafter \citetalias{2024ApJ...970..192H})\footnote{The data were obtained from the Mikulski Archive for Space Telescopes (MAST) at the Space Telescope Science Institute. The specific observations analyzed can be accessed via \dataset[doi:10.17909/26qj-g090]{https://doi.org/10.17909/26qj-g090}.}.
These datasets provide deep insights into various aspects of \omc{}, such as its spatial metallicity distribution \citep{2024ApJ...970..152N}, the age-metallicity relation \citep{2024ApJ...977...14C}, helium enrichment \citep{2025ApJ...984..162C}, 3D kinematics \citep{2025ApJ...983...95H}, and the discovery of a central intermediate-mass black hole \citep{2024Natur.631..285H}.

In this paper, we add new measurements of the equivalent width (EW) of the \nad{} absorption lines using MUSE spectra.
The spectra in the oMEGACat MUSE catalog \citepalias{2023ApJ...958....8N} were obtained from both Guaranteed Time Observations (GTO) and General Observer (GO) programs. They were fitted with the Phoenix spectral library \citep{2013A&A...553A...6H} using SPEXXY\footnote{\url{https://github.com/thusser/spexxy}} to derive basic stellar parameters such as \teff{}, $\log{g}$, and metallicity, as well as line-of-sight velocity. 
This analysis from \citetalias{2023ApJ...958....8N} is restricted to 342,797 main sequence (MS), subgiant branch (SGB), and red giant branch (RGB) stars. 
We refer to \cite{2018MNRAS.473.5591K, 2020A&A...635A.114H, 2021A&A...653L...8L} and \citetalias{2023ApJ...958....8N} for detailed descriptions.
For this study, we also include 60,054 additional MUSE spectra that were excluded from the oMEGACat MUSE catalog \citepalias{2023ApJ...958....8N}, which include horizontal branch (HB) stars. 
The GTO HB stars were previously analyzed by \cite{2023A&A...677A..86L}.

We use two photometric datasets in this study. 
The first is the HST/ACS catalog from \citealt{2010ApJ...710.1032A} (hereafter \citetalias{2010ApJ...710.1032A}), which was used as the reference catalog for PampelMUSE \citep{2013A&A...549A..71K} to extract MUSE spectra from GO observations.
The second is the oMEGACat HST catalog by \citetalias{2024ApJ...970..192H}, which provides precise photometry and proper motions from 20 years of HST archival data of \omc{}.
The photometric and astrometric information from these two catalogs is used for target selection in Section~\ref{subsec:data-target-selection} and plotting CMDs.
Additionally, the oMEGACat HST catalog \citepalias{2024ApJ...970..192H} is used to assess the crowding effect on our results due to its high completeness, and its photometry is used to evaluate the CMD correction based on the differential reddening derived in this work and previous studies.

In total, our data consists of 402,851 MUSE spectra with measurements of \teff{}, $\log g$, metallicity ([M/H]), and line-of-sight velocity (\vlos{}), along with multi-band photometry and astrometry.

\subsection{Combining the Datasets}
\label{subsec:data-others-combine}


Here we describe the procedures to add the photometry and astrometry from \citetalias{2010ApJ...710.1032A} and \citetalias{2024ApJ...970..192H} to our MUSE catalog.
Given the public oMEGACat MUSE catalog \citepalias{2023ApJ...958....8N} has already included \citetalias{2010ApJ...710.1032A} identifiers, we directly use these stellar IDs to add \citetalias{2010ApJ...710.1032A} information.
For spectra not included in the \citetalias{2023ApJ...958....8N} catalog and hence do not have \citetalias{2010ApJ...710.1032A} identifiers (e.g., newly added GTO spectra and horizontal branch stars), if \citetalias{2010ApJ...710.1032A} was used as the reference during spectra extraction, we perform cross-matching by finding the closest star in $(\alpha, \delta)$ with the same F625W magnitude. 
For spectra extracted using the \cite{2007AJ....133.1658S} catalog in F606W magnitudes, we cross-match with \citetalias{2010ApJ...710.1032A} by requiring $|\rm{F606W} - \rm{F625W}|<0.4$~mag to accommodate differences between the two filters and positional separation in $(\alpha, \delta)$ of less than 1~arcsec.
To incorporate \citetalias{2024ApJ...970..192H} photometry and astrometry, we cross-match \citetalias{2024ApJ...970..192H} with \citetalias{2010ApJ...710.1032A} by requiring a positional separation of less than 0.04~arcsec (1 HST WFC3/UVIS pixel) and magnitude differences of less than 0.5~mag in either F435W or F625W filters.
As a result, 400,945 MUSE spectra are successfully cross-matched with \citetalias{2010ApJ...710.1032A}, of which 384,978 also have information from \citetalias{2024ApJ...970..192H}.

\subsection{Target Selection}
\label{subsec:data-target-selection}

\begin{figure*}
\centering
\includegraphics[width=2\columnwidth]{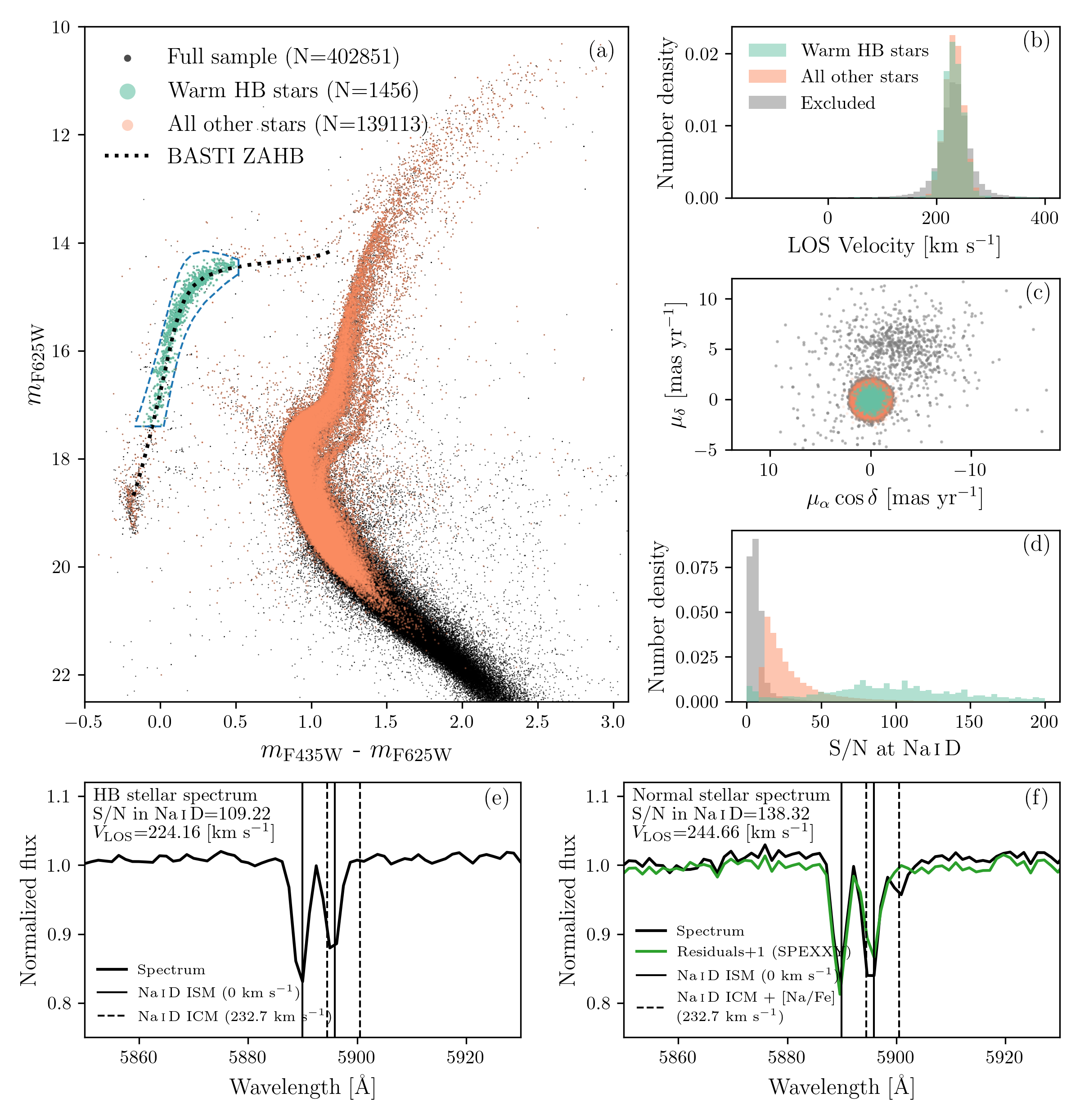}
\caption{
Data sample used in this study.
\textbf{Panel (a):} CMD of the data sample in Section~\ref{subsec:data-muse-hst} using \citetalias{2010ApJ...710.1032A} photometry. 
The blue region defines the selection of warm HB stars. 
The black dotted line shows the ZAHB track from the BaSTI model (see Section~\ref{subsubsec:data-target-selection-hb} for details).
\textbf{Panel (b):} \vlos{} distribution of the warm HB stars and all other stars. The gray histogram indicates stars excluded based on the selection criteria for the two samples.
\textbf{Panel (c):} Proper motion distribution of the same samples, with colors corresponding to panel (b). 
\textbf{Panel (d):} Re-estimated S/N distribution in the \nad{} region.
\textbf{Panel (e) and (f):} Example MUSE spectra (black line) of warm HB and all other stars, with the S/N and \vlos{} indicated in the top left corner.
The green lines show the residuals between the observed spectra and the best-fit SPEXXY model from \citetalias{2023ApJ...958....8N}.
Vertical lines are the positions of the \nad{} absorptions from foreground (solid, located at 0~\kms{}) and intracluster (dashed, at 232.7~\kms{}) atomic gas of \omc{}.
An additional contribution from the stellar photosphere, seen as the right-most absorption feature, is present in panel (f).
}
\label{fig:samplefig}
\end{figure*}


Here we describe the target selection procedures and the division of our sample. 
We first exclude MUSE spectra observed in AO-mode, because the wavelength around the \nad{} lines is blocked to avoid the strong laser light.
Next, we apply selection criteria based on line-of-sight velocity (\vlos{}) and proper motion ($\mu$) to identify \omc{} member stars, using the following equation:
\begin{equation}
\left\{\begin{array}{l}
179.9 < V_{\rm{LOS}} / \rm{km~s^{-1}} < 285.5  \\
\mu=\sqrt{\mu_\delta{ }^2+\mu_\alpha{ }^2 \cdot \cos ^2 \delta} < 3~\rm{mas}~\rm{yr}^{-1}
\end{array}\right.
\label{eqn:general-cutoff}
\end{equation}
where the \vlos{} range is within 3$\sigma$ of the mean velocity (232.7~\kms{}) and velocity dispersion (17.6~\kms{}) of \omc{} as reported by \cite{2018MNRAS.478.1520B}.
The proper motion cut-off applies only to stars having \citetalias{2024ApJ...970..192H} astrometry.
Then, we separate the remaining stars into two samples: warm HB stars and all other stars.
Fig.~\ref{fig:samplefig} shows the distribution of these two samples in the CMD as well as their distributions in \vlos{}, proper motion, and signal-to-noise ratio (S/N) in the \nad{} region (hereafter \snrnad{}). 
The bottom panels of Fig.~\ref{fig:samplefig} illustrate example MUSE spectra of these two samples of stars at the \nad{} region.
The criteria used to define these two samples are explained in the following subsections, and we define the \nadtwo{} line at $\lambda_{D_2} = 5889.950$~\AA{} and the \nadone{} line at $\lambda_{D_1} = 5895.924$~\AA{} throughout this work.

\subsubsection{Warm Horizontal Branch Stars}
\label{subsubsec:data-target-selection-hb}

The first sample consists of warm horizontal branch stars, most of which have higher effective temperatures (\teff{}) than MS, SGB and RGB stars.
Due to their higher \teff{}, these stars typically exhibit weaker metal lines from the stellar photosphere, particularly in the \nad{} region, as most elements are ionized.
This makes them an ideal sample for our study because we can assume that the \nad{} absorptions observed in HB spectra originate solely from the intracluster and foreground sodium atomic gas.
And we can use this sample to investigate both ISM and ICM distributions.

To select warm HB stars belonging to \omc{}, we define the blue region shown in Fig.~\ref{fig:samplefig}a.
This selection does three things: (1) it selects stars hotter than \teff{}~$\sim7500$~K; (2) it excludes the hottest HB stars (blue-hook stars) using a magnitude cut of $m_\mathrm{F625W}<17.4$~mag, because their lower luminosities result in poor spectral S/N, which can affect the analysis of intracluster gas in Section~\ref{sec:results-icm}; and (3) it includes stars that follow the expected horizontal as defined by two manually drawn boundaries.
We use the zero-age horizontal branch (ZAHB) track from the BaSTI $\alpha$-enhanced model \citep{2021ApJ...908..102P} as the expected horizontal branch (black dotted line in Fig.~\ref{fig:samplefig}a).
The BaSTI model has $\rm{[\alpha/Fe]=0.4}$, $\rm{[Fe/H]=-1.55}$~dex ($Z=0.000886$), and $Y=0.248$. The $m_\mathrm{F625W, BaSTI}$ magnitude is calculated using a distance modulus of $13.595$~mag \citep{2021ApJ...908L...5S} and $E(B-V)=0.13$~mag.
Note that the selection of the distance modulus and \ebv{} is arbitrary and only to match the BaSTI ZAHB with the observed HB stars. 
This blue selection region yields a sample of 1,456 warm HB stars.

The selected warm HB stars are highlighted in cyan in Fig.~\ref{fig:samplefig}.
Panel (c) shows the proper motion distribution of the selected warm HB stars and all other stars, along with the excluded stars from the full sample.
In panel (d), we show that most stars have signal-to-noise ratios (S/N) in the \nad{} region higher than 40~\perpixel{}.
An example warm HB spectrum in the \nad{} region is shown in panel (e) of Fig.~\ref{fig:samplefig}, where we marked the wavelength of \nad{} features contributed from intracluster and foreground atomic gas, respectively. 
We can see that the line strengths of foreground Na atomic gas are much stronger than those from the intracluster gas.

Although \citetalias{2023ApJ...958....8N} fitted some HB spectra using the SPEXXY code and derived \teff{}, the Phoenix templates they used from \cite{2013A&A...553A...6H} only have \teff{} up to $12000$~K, while many HB stars are known to be hotter \citep{2023A&A...677A..86L}.
Therefore, in this study, we re-estimate \teff{} using the color-\teff{} relation from the BaSTI ZAHB model and apply linear interpolation to determine the values.
We confirm that these \teff{} estimates are consistent with the spectroscopic measurements reported by \citet{2023A&A...677A..86L} and all our warm HB stars have \teff{} higher than 7450~K.

\subsubsection{All Other Stars}
\label{subsubsec:data-target-selection-non-hb}

The second sample consists of all other stars not highlighted in cyan in the CMD of Fig.~\ref{fig:samplefig}, primarily stars ranging from the MS to the RGB and a few hottest blue hook stars down the horizontal branch. 
Given that most of MS, SGB, and RGB stars are cooler than HB stars, they exhibit strong photospheric absorptions, particularly in the \nad{} region, due to neutral sodium in their atmospheres.
These absorption lines overlap with those of the intracluster atomic gas, making these spectra unsuitable for investigating the ICM.
However, we can still use them to study the foreground Na atomic gas, as the corresponding absorption-shown as the leftmost line in panel (f) of Fig.~\ref{fig:samplefig}-does not overlap with other features.

The MUSE spectra in this sample generally have worse data quality than warm HB stars due to their relatively lower luminosities (see panel (d) in Fig.~\ref{fig:samplefig}).
To exclude low-quality spectra, we make selection criteria based on \snrnad{} and PampelMUSE-reported spectral quality flag ($\rm{qflag_{PampelMUSE}}$) as follows:
\begin{equation}
\left\{\begin{array}{l}
\rm{S/N_{\ensuremath{\mathrm{Na\,\textsc{i}\,D}}}}>10~\rm{pix}^{-1}  \\
\rm{qflag_{PampelMUSE}=0}
\end{array}\right.
\label{eqn:nonhb-cutoff}
\end{equation}
The PampelMUSE quality flag removes spectra with an average S/N across the full wavelength below 10 \perpixel{}, negative mean flux, contribution from multiple sources, or source centroids located outside the data cube’s field of view. 
To ensure the data quality in \nad{} regions, we further remove spectra with \snrnad{} less than 10 \perpixel{}, where \snrnad{} is estimated with detailed descriptions in Section~\ref{subsec:data-reestimate-snr}.
We also remove stars with failed SPEXXY fits which do not have residuals.
After applying these criteria, we end up with 139,113 stars.

We show the locations of this sample on the CMD along with \vlos{}, proper motion, and \snrnad{} distributions in orange in Fig.~\ref{fig:samplefig}.
Panel (f) demonstrates an example of the spectrum (black) and SPEXXY residuals (green) between the observed data and the best-fit SPEXXY model from \citetalias{2023ApJ...958....8N}, with a constant offset added for visibility.
We can see a more significant \nad{} absorption at \omc{} systemic velocity (right-most line) than the case for the warm HB spectra in panel (e), and it becomes less prominent in SPEXXY residuals.
This indicates that the \nad{} feature was reproduced by the best-fitting Phoenix spectrum and not present in the residuals.
In addition, the \nadtwo{} absorption at the foreground velocity (left-most line) remains strong.
Based on these findings, we use the SPEXXY residuals in the following sections to investigate the foreground atomic gas.
Compared to the warm HB spectra sample, which has individual stellar measurements, this sample can provide a 2D distribution map of foreground atomic gas. 
We do not use this sample to investigate the intracluster gas distribution.
This is because the [Na/Fe] ratio in the Phoenix templates used by SPEXXY is fixed to the solar value, and the star-to-star variation is not considered. 
However, [Na/Fe] variations in \omc{} are non-negligible due to the presence of multiple populations \citep[e.g.,][]{2010ApJ...722.1373J}.
Therefore, the sodium absorption due to stellar photosphere can remain in the SPEXXY residuals.



\subsection{Spectral Normalization and Re-estimation of S/N in \texorpdfstring{\nad}{} Region}
\label{subsec:data-reestimate-snr}

Here we describe the data preprocessing procedures conducted before fitting the spectra to measure the \nad{} EW.
First, we perform a pseudo-continuum normalization on the full wavelength range of the MUSE spectra by dividing the observed flux by a Gaussian kernel with a width of 50~\AA{}. 
This method was previously applied to LAMOST and MUSE observations \citep[e.g.,][]{2019ApJS..245...34X, 2022MNRAS.514.1034W} and can effectively remove the continuum.
Next, given that the PampelMUSE-provided S/N tend to be under-estimated \citep[e.g.,][]{2022MNRAS.514.1034W}, we re-estimate the S/N to ensure an accurate determination of the \nad{} EWs and the uncertainties.
For each star, we reapply the pseudo-continuum normalization in the wavelength regions $5850<\lambda/\angstrom<5870$ and $5905<\lambda/\angstrom<5930$, as shown in gray regions of Fig.~\ref{fig:fitting-example}, to reduce small-scale spectral wiggles in the \nad{} region. 
These wavelength regions are selected to avoid \nad{} and $\mathrm{He\,\textsc{i}}$ absorption (5875.6~\AA{}) features where the latter are stronger in HB stars with larger \teff{} \citep{2014A&A...563A..13G}.
We then rescale the flux errors ($e_{obs}$) using the factor defined as follows:
\begin{equation}
\text{factor} = \frac{\mathrm{std}(f_{obs}-f_{c})}{\mathrm{mean}(e_{obs})},
\label{eqn:snr-reestimate}
\end{equation}
where $f_{c}$ is the mean flux of the continuum, and $f_{obs}$ is the observed flux. 
After rescaling the flux errors, the \snrnad{} is recalculated in the wavelength range of $5850<\lambda/\angstrom<5930$.
In this study, when referring to flux uncertainty or S/N, we are always referring to these rescaled flux errors.


\section{Methods}
\label{sec:methods}

\subsection{Measuring the \texorpdfstring{\nad}{} Equivalent Width Using Markov Chain Monte Carlo}
\label{subsec:methods-ew}

\begin{figure}
    \centering
    \includegraphics[width=1\linewidth]{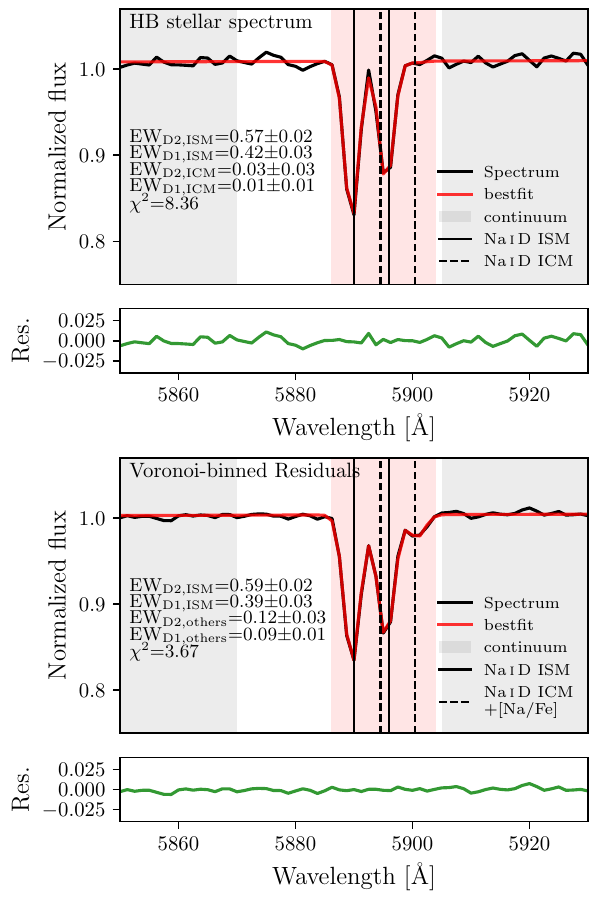}
    \caption{
    Demonstration of spectral fitting to measure the EWs of \nad{}, as described in Section~\ref{subsec:methods-ew}.
    The top panel shows an example spectrum of a warm HB star, while the bottom panel shows a Voronoi-binned residual spectrum from all other stars.
    The equivalent widths (EWs) of the \nad{} lines, in units of \AA{}, are labeled in each panel.
    The red-shaded region indicates the wavelength range used for fitting.
    The gray-shaded regions are used to estimate the continuum ($f_{c}$) in Eq.~\ref{eqn:nad-fit}. 
    Fitting residuals are shown in green as a separate sub-panel.
    The warm HB spectrum is expected to have minimal photospheric contribution to \nad{} and thus we name them as ``ISM'' and ``ICM''.
    For all other stars that are mostly cooler, contamination from the stellar photosphere could remain in the SPEXXY residuals.
    Therefore, we label the absorptions at \omc{} systemic velocity as ``others''. 
    Note that the D2 line is at a shorter wavelength than the D1 line for both the ISM and ICM/others components.
    }
    \label{fig:fitting-example}
\end{figure}

The \nad{} absorptions contributed by intracluster and foreground atomic gas are modeled in the wavelength region of $5886<\lambda/\angstrom<5904$ as the summation of a mean continuum $f_c$ and four Gaussian profiles.
The use of Gaussian functions follows the approach described in \cite{2009MNRAS.399..195V}.
Then the \nad{} absorptions are described by the following equation:
\begin{align}
\begin{split}
    f(\lambda) = f_{c} \times [1 &+ A_{\rm{ISM,D_2}} \times e^\frac{{-(\lambda - \lambda_{\rm{ISM,D_2}})^2}}{2\sigma_{\rm{ISM}}^2} \\
    &+ A_{\rm{\rm{ISM},D_1}} \times e^\frac{{-(\lambda - \lambda_{\rm{ISM,D_1}})^2}}{2\sigma_{\rm{ISM}}^2} \\
    &+ A_{\rm{ICM,D_2}} \times e^\frac{{-(\lambda - \lambda_{\rm{ICM,D_2}})^2}}{2\sigma_{\rm{ICM}}^2} \\
    &+ A_{\rm{ICM,D_1}} \times e^\frac{{-(\lambda - \lambda_{\rm{ICM,D_1}})^2}}{2\sigma_{\rm{ICM}}^2}],
\label{eqn:nad-fit}
\end{split}
\end{align}
where $f_c$ is derived in the same way as Section~\ref{subsec:data-reestimate-snr}.
$\lambda_{\rm{ISM}}$ and $\lambda_{\rm{ICM}}$ are the central wavelength of \nad{} lines contributed by foreground and intracluster atomic gas, respectively.
The wavelengths of \nad{} from foreground atomic gas (ISM terms in the above equation) are constrained to an equivalent \vlos{} range of ($-$50, 50)~\kms{}, consistent with the velocity distribution of foreground stars in Gaia DR3 \citep{2023A&A...674A...1G}. For the intracluster atomic gas (ICM terms in Eq.~\ref{eqn:nad-fit}), the \nad{} absorption is constrained to ($-$20, 20)~\kms{} relative to the systemic velocity of \omc{} (232.7 \kms{}), corresponding to the $\sim$1$\sigma$ stellar velocity dispersion.  
We are therefore assuming that the intracluster gas is co-moving with the cluster.
The line widths $\sigma_{\rm{ISM}}$ and $\sigma_{\rm{ICM}}$ are constrained to ranges of (0.8, 1.8)~\AA{} and (0.8, 2.5)~\AA{}, respectively, and they are forced to be the same for $\rm{D_2}$ and $\rm{D_1}$ lines.
The amplitudes ($A_{\rm{ISM, D_2}}$, $A_{\rm{ISM, D_1}}$) and ($A_{\rm{ICM, D_2}}$, $A_{\rm{ICM, D_1}}$) are independently constrained to ($-$1, 1) and ($-$0.5, 0.5), respectively, allowing for variability between these components.
These parameter constraints were determined based on tests to ensure that fits were not focusing on weak, spurious features from single outlier wavelength pixels.  

During the fitting, we first obtain the initial parameter guesses for Eq.~\ref{eqn:nad-fit} using \texttt{scipy.curve\_fit}.
Next, we apply Markov chain Monte Carlo (MCMC) to estimate the parameters and uncertainties.
This is done by using the \textsc{emcee} code \citep{2013PASP..125..306F} with 50 walkers and 1500 iterations.
The parameters and uncertainties are derived from the median and ($16^{th}$, $84^{th}$) percentiles of posterior distributions.
We also provide $\chi^2=\Sigma \frac{[f_{obs}-f(\lambda)]^2}{e_{obs}^2}$ to quantify the fitting quality.
Because the spectrum is continuum-normalized, the EW of each \nad{} line is then calculated by using the Gaussian integration equation, $\rm{EW} = |A| \sqrt{2\pi\sigma^2}$, with uncertainties determined by error propagation.
We also derive the \vlos{} of both \nad{} contributions using the Doppler equation based on the wavelength shifts between the Gaussian centroids and their expected wavelengths.
We use \ewismdtwo{} and \ewicmdone{} to study the foreground and intracluster Na atomic gas, respectively, as \ewismdone{} and \ewicmdtwo{} (the middle line in the \nad{} region) overlap and can not be separated.

For warm HB stars, the fitting is performed directly on each individual MUSE spectrum.
For all other stars, we first divide the field into a regular grid with a cell size of $7.5''\times7.5''$. 
Each cell contains a median of 16 stars.
For each cell, we compute the S/N-weighted average SPEXXY residual and its average \snrnad{}. 
Next, we apply Voronoi binning \citep{2009arXiv0912.1303C} to these gridded residual spectra, using the average \snrnad{} of each cell to define the Voronoi bins, with a target \snrnad{} of 300~\perpixel{}.
The \nad{} EWs are then measured from these Voronoi-binned residual spectra. 
Note that the gridding has two purposes: (1) it is required by the Voronoi-binning code, and (2) it enforces a minimum bin size, ensuring that no Voronoi bin is contributed by a single high-S/N spectrum.
We obtain 1662 Voronoi bins with sizes ranging from 1 to 38 cells and a median of 4 cells. 
Each Voronoi bin contains between 9 and 405 stars, with a median of 76 stars. 
The use of SPEXXY residuals rather than original spectra is to minimize the influence of stellar photospheric absorptions in the \nad{} region, which could otherwise bias the determination of the mean continuum flux $f_c$ and in the end affect the EW measurements. 
An example of the fitting procedure applied to warm HB and Voronoi-binned residual spectra is shown in Fig.~\ref{fig:fitting-example}.
This figure indicates that our fitting algorithm works for both types of spectra.

\subsection{Linking \texorpdfstring{\nad}{} Equivalent Width to Extinction}
\label{subsec:methods-ebv}

Despite the \nad{} lines tracing Na atomic gas, previous work has found a clear correspondence between the \nad{} line strengths and general foreground dust extinction of a source \citep{2012MNRAS.426.1465P, 2015MNRAS.452..511M}.  
Therefore, we can convert our derived EWs of the \nad{} lines to \ebv{} using existing empirical relations. 
Specifically, we use \ewismdtwo{} (the left-most \nad{} line in Fig.~\ref{fig:fitting-example}) to determine the foreground ISM reddening, applying the empirical relation from \cite{2012MNRAS.426.1465P} as follows:
\begin{equation}
    E(B-V) = 10^{(2.16 \times \rm{EW}_{\rm{D2, ISM}} - 1.91\pm0.15)}.
\label{eqn:ebvewism}
\end{equation}
This relation was derived by fitting \nad{} EWs measured from both low- and high-resolution quasar (QSO) spectra from SDSS and Keck/HIRES, respectively, with reddening from the Schlegel, Finkbeiner \& Davis map \citep[][hereafter SFD]{1998ApJ...500..525S}. 
The typical relative uncertainty of \ebv{} from this equation is 35\%.
This relation is fitted for \ebv{} up to 3~mag.
Most of the data used in the fit have \ebv{}~$<0.09$~mag (EW~$<0.4$~\AA{}), with only four data points covering the \ebv{} range of $0.09<$~\ebv/mag$~<2.92$ ($0.4<$~EW~$<1.1$~\AA{}).
Notebly, \cite{2015MNRAS.452..511M} presented a similar relation to Eq.~\ref{eqn:ebvewism} and suggested that the relation from \cite{2012MNRAS.426.1465P} may be affected by stellar contamination in their sample. However, since the \cite{2015MNRAS.452..511M} relation is restricted to $E(B-V) < 0.08$, we still adopt the relation from \cite{2012MNRAS.426.1465P} throughout this study. We also perform our calculations using the \cite{2015MNRAS.452..511M} relation and found that our results and conclusions remain unchanged.

Considering the applicable range of Eq.~\ref{eqn:ebvewism} and its uncertainty, in the following sections, we mainly use \ewismdtwo{} to study the foreground distribution without converting to \ebv{}, and only use Eq.~\ref{eqn:ebvewism} to obtain an average \ebv{} and differential reddening.
We apply a correction factor of 0.86 to the reddening values as suggested by \cite{2012MNRAS.426.1465P} to account for the over-estimation reported by \cite{2010ApJ...725.1175S, 2011ApJ...737..103S} in the SFD dust distribution.
For the ICM, given the \ewicmdone{} measurement of a warm HB spectrum in Fig.~\ref{fig:fitting-example} (right-most line) is much smaller with large uncertainty, in this study, we focus only on assessing the significance of this absorption feature.



As discussed in Section~\ref{subsubsec:data-target-selection-hb}, the warm HB sample is used for both ISM and ICM investigations. 
For all other stars, although the contamination due to stellar photosphere can be partially alleviated using the SPEXXY residual spectra (as discussed in Section~\ref{subsubsec:data-target-selection-non-hb}), it cannot be fully removed. 
The Voronoi-binned residuals in the bottom panel of Fig.~\ref{fig:fitting-example} demonstrate that the right-most \nad{} absorption remains significant.


\section{Mapping the Sodium Absorption from the Interstellar Medium} 
\label{sec:results-ism}

In this section, we present the distribution of the \nad{} absorption from the foreground atomic gas.
We first illustrate the spatial distribution of \ewismdtwo{} in $(\alpha, \delta)$.
Next, we use these measurements to calculate the foreground average \ebv{} and the differential reddening across the face of \omc{}.

\subsection{The Spatial Distribution of the \texorpdfstring{\nad}{} Equivalent Width}
\label{subsec:results-ism-2d}

\begin{figure*}
    \centering
    \includegraphics[width=1\linewidth]{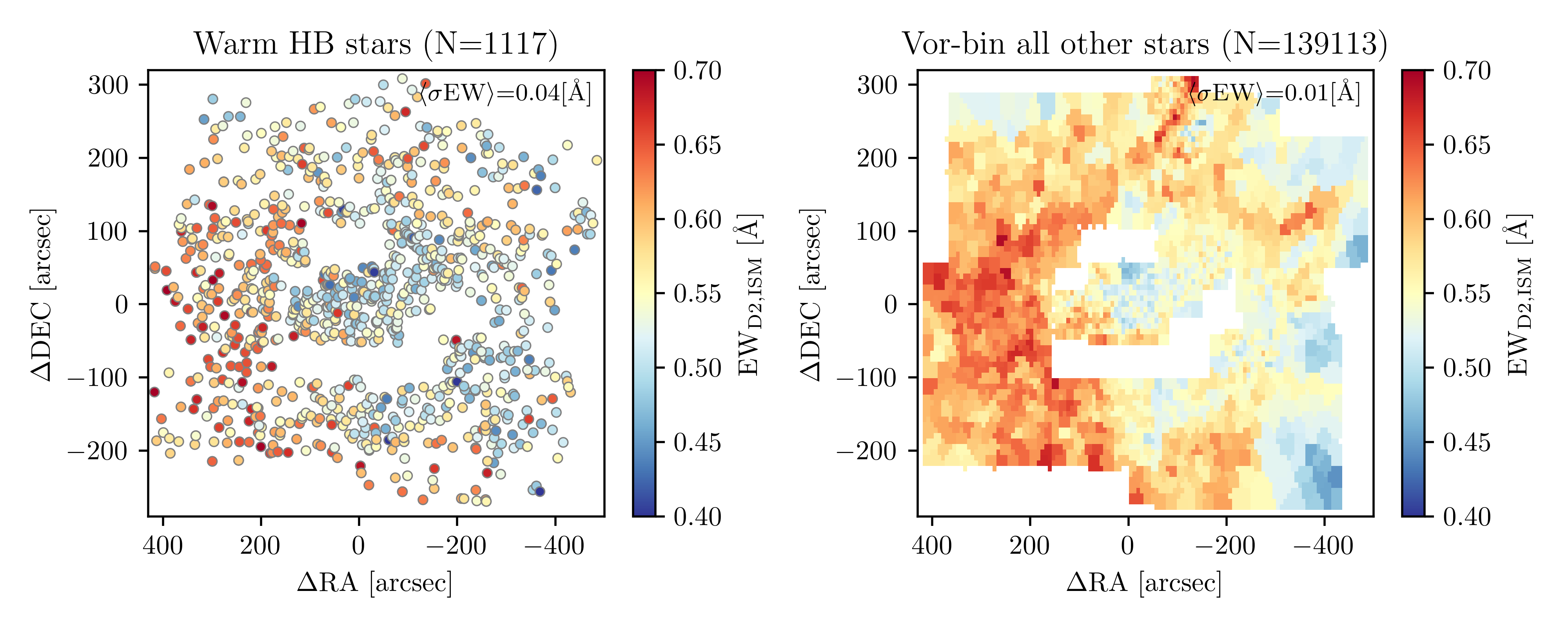}
    \caption{
    Spatial distribution of \ewismdtwo{} from the foreground Na atomic gas.
    The left panel shows individual measurements from the original spectra of warm HB stars.
    The right panel shows measurements from Voronoi-binned SPEXXY residual spectra \citepalias{2023ApJ...958....8N} of all other stars.
    The number of stars used for each panel is noted in the subtitles.
    The x- and y-axes indicate coordinates in $(\alpha, \delta)$ relative to the center of \omc{}.
    The two blank regions near the center of \omc{} are due to the lack of non-AO observations.
    Both warm HB and all other stars demonstrate clear spatial substructures of foreground Na atomic gas, and they are consistent with the differential reddening distribution from previous studies (see Section~\ref{subsec:discuss-comparison} for detailed discussion).
    The individual warm HB \ewismdtwo{} measurements and the Voronoi-binned \ewismdtwo{} map with uncertainties are available as a machine-readable table (columns listed in Table.~\ref{tab:ewd2ism_table}) and a FITS image, respectively.
    }
    \label{fig:ew-ism-distribution}
\end{figure*}

\input{ewd2ism_table}

\begin{figure}
    \centering
    \includegraphics[width=0.9\linewidth]{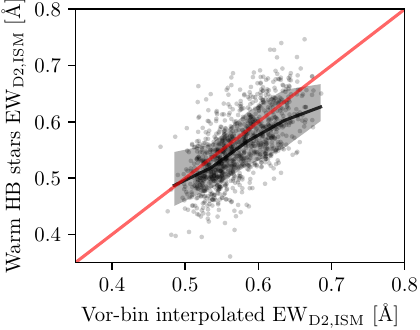}
    \caption{
    Comparison of \ewismdtwo{} measurements between warm HB stars and all other stars as shown in Fig.~\ref{fig:ew-ism-distribution}.
    The gray dots represent individual warm HB stars.
    The y-axis represents \ewismdtwo{} measured from individual warm HB spectra, and the x-axis corresponds to \ewismdtwo{} values interpolated from the Voronoi-binned maps of all other stars, using the coordinates of the warm HB stars.
    The red line indicates a one-to-one relation for reference.
    The black line and gray-shaded region represent the median and the ($16^{th}$, $84^{th}$) percentiles of the comparison, respectively.
    This figure indicates that although warm HB stars and all other stars show a similar structure in their spatial distributions, their one-to-one comparison reveals a systematic offset.
    }
    \label{fig:ew-ism-compare}
\end{figure}

We present in Fig.~\ref{fig:ew-ism-distribution} the spatial distributions of \ewismdtwo{} contributed by the foreground atomic gas.
The left panel shows individual \ewismdtwo{} measurements from warm HB spectra, and the right panel shows results from Voronoi-binned SPEXXY residual spectra of all other stars.
The distributions are color-coded by \ewismdtwo{}.
The x- and y-axis represent coordinates in $(\alpha, \delta)$ relative to the center of \omc{} at $(13^{\mathrm{h}}26^{\mathrm{m}}47.24^{\mathrm{s}}, -47^\circ 28' 46.45'')$ from \citetalias{2010ApJ...710.1032A}.
The two blank regions near the center are due to the lack of non-AO observations.
To ensure measurement accuracy, we apply the cut-off on the warm HB stars as follows:
\begin{equation}
\left\{\begin{array}{l}
\rm{S/N_{\ensuremath{\mathrm{Na\,\textsc{i}\,D}}}}>40~\rm{pix}^{-1},  \\
\chi^2<15,
\end{array}\right.
\label{eqn:hbew-cutoff}
\end{equation}
where $\chi^2$ represents the fitting quality as defined in Section~\ref{subsec:methods-ew}.
This cut-off is based on the changes of \ewismdtwo{} uncertainties and $\chi^2$ with \snrnad{}, resulting in 1,117 stars.
The median \ewismdtwo{} uncertainty of the remaining warm HB stars is 0.04~\AA{}.
For results from the Voronoi-binned residual spectra, no selection cut-off is applied because their \snrnad{} are about 300~\perpixel{}, which has a smaller average \ewismdtwo{} uncertainty with a median of 0.01~\AA{}.
The individual warm HB \ewismdtwo{} measurements with uncertainties and BASTI-ZAHB track-inferred \teff{} are provided as a machine-readable table, summarized in Table~\ref{tab:ewd2ism_table}. 
The Voronoi-binned \ewismdtwo{} distribution is also available in machine-readable FITS format.

In Fig.~\ref{fig:ew-ism-distribution}, both datasets show that \ewismdtwo{} values are higher at positive $\Delta \rm{RA}$ than those at negative $\Delta \rm{RA}$.
The inner region ($|\Delta \rm{RA}|$ and $|\Delta \rm{DEC}|$ less than 100~arcsec) exhibits smaller \ewismdtwo{} than the outer region.
These spatial substructures are similar to the differential reddening distributions in previous studies (e.g., \citealt{2017ApJ...842....7B, 2024AA...686A.283P} and \citetalias{2024ApJ...970..192H}).
We discuss the comparison to literature differential reddening estimates in detail in Section~\ref{subsec:discuss-comparison}.
In addition, the similar substructures observed between warm HB stars and all other stars demonstrate the consistency of our results across the two samples.

We show in Fig.~\ref{fig:ew-ism-compare} the comparison of \ewismdtwo{} measurements from warm HB stars and all other stars.
The values on the x-axis are obtained by linearly interpolating the \ewismdtwo{} values from the spatial distribution of Voronoi-binned results at the positions of the warm HB stars.
We find that the \ewismdtwo{} values from warm HB and all other stars are qualitatively consistent.

However, the median trend (gray) exhibits a systematic offset compared to the one-to-one line (red), with the Voronoi binned residual spectra showing somewhat larger \ewismdtwo{} than the hot star sample.
This offset is likely caused by the adoption of SPEXXY residual spectra and Voronoi binning.
For example, some absorption features could remain in the SPEXXY residuals because the Phoenix spectra have solar-scaled abundances, so they can still affect the continuum determination and hence affect the EW estimation.
Moreover, potential artifacts may be present in the residuals because the SPEXXY best-fit continuum may not perfectly match the observed spectrum.
Furthermore, although the Phoenix templates have been degraded to match the MUSE spectral resolution (see details in \citetalias{2023ApJ...958....8N}), the adopted line-spread function is an average, and its variation between different MUSE IFUs is not considered \citep[e.g., ][]{2016A&A...588A.148H}. 
Consequently, the absorption profiles may differ between the MUSE spectra and the Phoenix library, and cause the profile shapes of Voronoi-binned residual spectra to become non-Gaussian.
This can be verified by comparing the width of the Gaussian profile ($\sigma_{\rm{ISM}}$) from residual and original spectra, where we find that $\sigma_{\rm{ISM}}$ from original spectra is smaller than that from residual spectra of all other stars.
We also compare our warm HB measurements to those derived from the Voronoi-binned original spectra of all other stars and found that the resulting median trend is in excellent agreement.
Since the results from warm HB stars are measured directly from the original MUSE spectra and are less affected by stellar photospheric absorption, we use these as our primary results in the remainder of the paper.

\subsection{Estimation of \ebv{} and Differential Reddening}
\label{subsec:results-ism-dr}

\input{reddening_table}

We use Eq.~\ref{eqn:ebvewism} in Section~\ref{subsec:methods-ebv} to convert \ewismdtwo{} measurements of both warm HB stars and all other stars into the foreground reddening, \ebv{}.
The median uncertainty of \ebv{} from warm HB stars converted using Eq.~\ref{eqn:ebvewism} is 0.066~mag when considering the systematic uncertainties in the equation.
However, in deriving the differential reddening, we do not include this systematic uncertainty on the conversion of EW to \ebv{}, since we are interested in the spread in measured values.
Instead, we calculate the \ebv{} uncertainties just from the EW errors by taking half the difference between \ebv{} values derived using $\rm{EW} + \sigma{\rm{EW}}$ and $\rm{EW} - \sigma{\rm{EW}}$.
Using this approach, the obtained median \ebv{} uncertainty of individual warm HB stars is 0.033~mag.
We then estimate the intrinsic \ebv{} distribution by assuming it is a Gaussian; we use the methods outlined in \citet{1993ASPC...50..357P} to deconvolve the reddening measurements while accounting for the individual measurement errors \citep[for details on our implementation, see Section 5.2 of][]{2024ApJ...977...14C}.
The derived intrinsic Gaussian mean then gives us the average \ebv{} in our field-of-view, while the inferred dispersion quantifies the differential reddening.

After the above calculations, the intrinsic average \ebv{} of warm HB stars is $0.153 \pm 0.003$~mag, with the error estimated through 300 bootstrap resamplings of the individual warm HB stars (and not including any systematic uncertainty on the conversion between EW and \ebv{}).
From the Voronoi-binned residual spectra of all other stars, we obtain a larger average \ebv{} of $0.182 \pm 0.002$~mag; as noted above, this difference is due to the offsets of \ewismdtwo{} measurements as seen in Fig.~\ref{fig:ew-ism-compare}.
We consider the warm HB star estimate to be more reliable, and thus adopt a mean \ebv{} of 0.153~mag.

Our best estimate mean \ebv{} value from the warm HB stars is higher than 0.12~mag reported by \cite{2010arXiv1012.3224H}, which was compiled from \citet{1988PASP..100..545R, 1985IAUS..113..541W, 1985ApJ...293..424Z} using data located at radii $\gtrsim 13.5’$ from the center of \omc{}, and also higher than the $0.08\pm0.02$~mag reported by \citep{2009MNRAS.394..831M} over a field of view with a radius of $0.5^{\circ}$.
A more recent photometric reddening estimate is also available using the oMEGACat dataset \citep{2024ApJ...977...14C};
they used the oMEGACat metallicity estimates to compare the CMD of low-metallicity stars \omc{} with two low-extinction globular clusters NGC~4147 and NGC~7089, and found that an \ebv{} of 0.185~mag is required to align the MSTO region of \omc{} with the other two clusters.  
They quantify systematics due to differing helium abundances and suggest a maximum error in this value of 0.04~mag, thus their value is in reasonable agreement with the completely independent estimate we get from \nad{} absorption.
Although previously published reddening values are smaller than our estimates and that from \cite{2024ApJ...977...14C}, this discrepancy could be because the measurements are not obtained over the same spatial region.
Indeed, the differential reddening maps of \cite{2024AA...686A.283P} show a higher reddening in the central regions than at larger radii, possibly explaining the difference between the \cite{2010arXiv1012.3224H} and the two oMEGACat-based values.  

In this work, differential reddening is derived as the intrinsic spatial spread of reddening values by removing measurement errors.
Our method assumes this spread is described by a Gaussian.  
Based on the warm HB stars and all other stars we find intrinsic spreads with a Gaussian width of $0.026\pm0.003$~mag and $0.035\pm0.001$~mag, respectively; the errors are estimated in the same bootstrap strategy as average \ebv{}.
We also use our warm HB stars and adopt the approach commonly employed in previous studies \citep[e.g.,][]{2017ApJ...842....7B, 2022MNRAS.517.5687J, 2024AA...686A.283P, 2024ApJ...970..192H, 2024A&A...690A..37S} to estimate differential reddening from photometry. 
For each warm HB star, we calculate its color displacement relative to the median fiducial HB line in the CMD ($\mathrm{F814W}$ vs. $\mathrm{F275W}–\mathrm{F814W}$), and convert this displacement into \ebv{}. 
We then apply the error deconvolution method of \citet{1993ASPC...50..357P} (also used in \citealt{2024ApJ...977...14C}) to derive the intrinsic width of the reddening distribution, which we interpret as the photometric differential reddening.
Using this approach, the photometric differential reddening is estimated to be $0.018\pm0.003$~mag; this estimate is an upper limit as it assumes that the intrinsic width of these stars on the CMD is zero.  Thus immediately, we can see there is a tension between this result and the \nad{} intrinsic width.  
To further compare our differential reddening measurements, we summarize our results alongside literature values, measurement methods, and FoV in Table~\ref{tab:reddening_dr}.
Our \nad{} differential reddening measurement is also much higher than literature values obtained using photometric methods except \cite{2005ApJ...634L..69C}.
We discuss this mismatch in detail in Section~\ref{subsec:discuss-dr-mismatch}.

\section{Investigating the Sodium Absorption due to the Intracluster Medium}
\label{sec:results-icm}

\begin{figure*}
    \centering
    \includegraphics[width=1\linewidth]{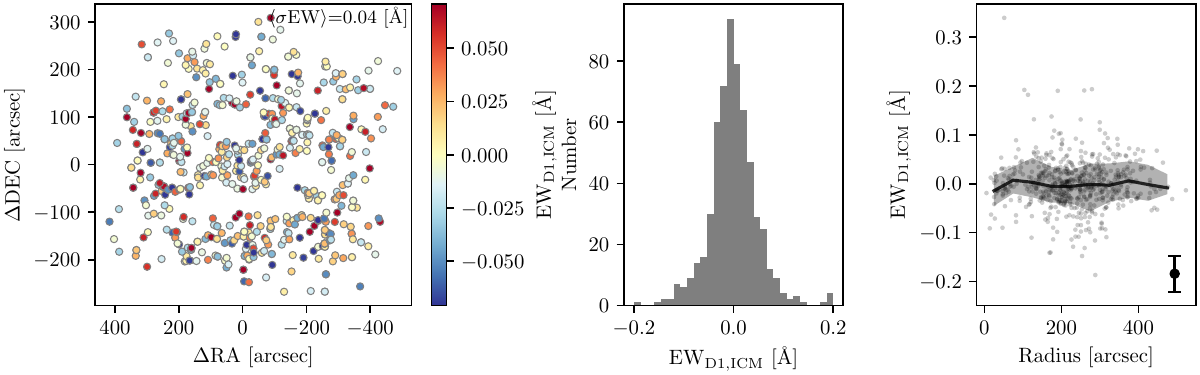}
    \caption{
     \ewicmdone{} distribution from the intracluster Na atomic gas of \omc{}, which is measured from individual warm HB spectra.
    \textbf{Left panel:} Spatial distribution of warm HB stars color-coded by \ewicmdone{}.
    \textbf{Middle panel:} Histogram of the \ewicmdone{} measurements in the left panel.
    \textbf{Right panel:} Radial distribution of \ewicmdone{} from the center of \omc{}, where the black line indicates the median of individual measurements, and the gray regions represent the $16^{th}$ and $84^{th}$ percentiles for each bin. The \ewicmdone{} median uncertainty is also shown in the bottom right corner in black.
    This figure shows no significant spatial substructure and radial gradient of Na atomic gas inside \omc{}.
    The individual warm HB \ewicmdone{} measurements used in this figure are available as a machine-readable table with columns listed in Table.~\ref{tab:ewd2ism_table}.
    }
    \label{fig:fitting-icm-individual}
\end{figure*}

\begin{figure}
    \centering
    \includegraphics[width=1\linewidth]{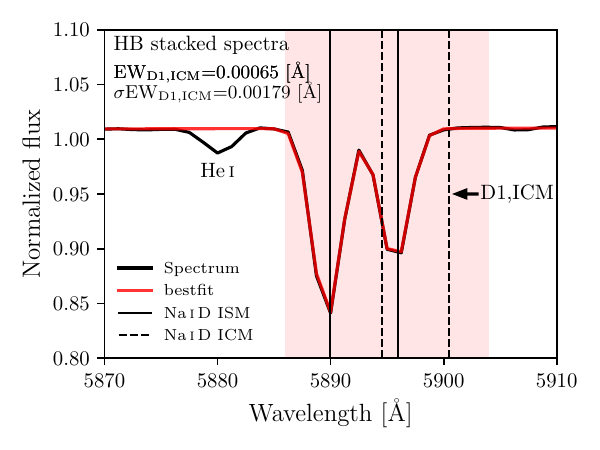}
    \caption{
    Spectral fitting of the stacked spectrum from all warm HB stars with S/N~$>40$~\perpixel{} and \teff{}~$>10000$~K.
    The definitions of the colors of each line and region are consistent with those in Fig.~\ref{fig:fitting-example}.
    This figure primarily investigates the presence of \nad{} absorption attributed to the intracluster gas of \omc{}.
    Based on the EW of D1 line (right-most absorption), no significant amount of intracluster Na atomic gas is detected.
    }
    \label{fig:fitting-icm-stacked}
\end{figure}

In this section, we investigate whether any signatures of intracluster sodium atomic gas in \omc{} can be detected using the warm HB spectra.
We focus only on the EW measurement of the D1 line from the intracluster gas, i.e., \ewicmdone{}, as it is not contaminated by foreground absorption, although the D1 line is weaker.
We use individual warm HB spectrum measurements of \ewicmdone{} to explore the spatial and radial distributions of intracluster atomic gas.
We also perform the fitting on a stacked spectrum of all warm HB stars with very high \snrnad{} and determine whether any significant \ewicmdone{} feature can be detected.
To minimize contamination from stellar photospheric absorption and ensure high spectral quality, we selected warm HB stars with
\teff{}~$>10000$~K and \snrnad{}~$>40$~\perpixel{}. This selection resulted in a sample of 595 stars. 
The individual warm HB \ewicmdone{} measurements with uncertainties are also included in the machine-readable table with columns summarized in Table~\ref{tab:ewd2ism_table}.
We note that we performed tests of this sample of stars as a function of their local background and number of neighbors to assess whether nearby stars could contaminate this measurement and found no significant impact.

Fig.~\ref{fig:fitting-icm-individual} shows the distribution of individual \ewicmdone{} estimates.
From the left and middle panels, we find no clear spatial structure in the distribution of sodium atomic gas from ICM, and our measurements demonstrate a comparable number of positive and negative values.
The mean and standard error of the mean (SEM) of \ewicmdone{} is $-0.0004\pm0.0020$~\AA{}. 
In the right panel of Fig.~\ref{fig:fitting-icm-individual}, we show the radial distribution of \ewicmdone{} by taking the median of each radius bin and find no significant radial gradient of the distribution.
Although some data points are far from zero, we find that the standard deviation of \ewicmdone{}$/\sigma$\ewicmdone{} is 1.05, suggesting the scatter is consistent with the errors on the individual measurements.
Therefore, these results suggest that the absorption of \nadone{} line from intracluster atomic gas is very weak. 

We also measure the \ewicmdone{} on a stacked warm HB spectrum to investigate whether our non-detection of intracluster atomic gas is due to the low \snrnad{} of individual warm HB spectra.
The stacked spectrum has a \snrnad{} of $\sim2000$~\perpixel{} and the fitting result is shown in Fig.~\ref{fig:fitting-icm-stacked}.
We can see that no significant feature is found at \nadone{} from intracluster gas.
The estimated \ewicmdone{} is $0.0007\pm0.0018$~\AA{}, fully consistent with the absence of intracluster atomic sodium gas in \omc{}. 
In the following sections, we adopt these values from the stacked spectra as our intracluster sodium measurement, and the uncertainty is used to calculate the upper limit of intracluster atomic gas density in Section~\ref{subsec:discuss-icmgasdensity}.


\section{Discussion} 
\label{sec:disscuss}

\subsection{Upper Limit of Gas Density in  \texorpdfstring{\omc}{}}
\label{subsec:discuss-icmgasdensity}

Although we do not detect a significant presence of Na atomic gas in \omc{}, our \nad{} measurements still allow us to place an upper limit on the atomic gas density within the cluster. 
In this section, we first use stellar parameters from the oMEGACat dataset to estimate the expected mass loss from \omc{} stars. 
We then use our \nad{} results to infer upper limits on the intracluster gas density and the timescale required for this gas to accumulate via mass loss. 
Finally, we explore potential reasons for the lack of ICM.

\subsubsection{Mass Loss Rate from Giant Stars in \texorpdfstring{\omc}{}}
\label{subsubsec:discuss-icmgasdensity-massloss}

The mass loss of stars provides a constant source of gas in the cluster, primarily contributed by bright giant stars.
Therefore, we select giant stars, starting from the end of the SGB phase, using the oMEGACat MUSE catalog \citepalias{2023ApJ...958....8N}.
After applying the \omc{} membership criteria defined in Eq.~\ref{eqn:general-cutoff}, we further apply the following cut-offs to identify giant stars:
\begin{equation}
    \left\{\begin{array}{l}
    m_{\mathrm{F625W}}<18~\rm{mag}, \\
    1<m_{\mathrm{F435W}}-m_{\mathrm{F625W}}<3, \\
    \teff{}<6500~\rm{K}, \\
    \log{g}<3.7~\rm{dex},  \\
    \end{array}\right.
\label{eqn:rgb-selection}
\end{equation}
where the photometry is from \citetalias{2010ApJ...710.1032A}, \teff{} and $\log{g}$ are provided by \citetalias{2023ApJ...958....8N}.
This selection results in a total of $N_{\rm{giant}}=13,130$ stars.
Because all selected stars are brighter than 18~mag and 95\% have MUSE S/N greater than 11.7~\perpixel{}, this giant sample should be nearly complete within the half-light radius of \omc{}.

To calculate the total mass loss rate, we first estimate the average stellar mass loss rates for both helium-rich ($\dot{M}_{\rm{He,R}}$) and primordial-helium ($\dot{M}_{\rm{He,P}}$) populations.
This is done by taking the stellar mass difference between RGB and HB phases and dividing it by the RGB lifetime: $\dot{M}=\frac{M_{\rm{RGB}}-M_{\rm{HB}}}{\tau}$.
The RGB stellar masses are obtained by interpolating the $\log g$–$M$ relation at $\log g=3.7$ from helium-rich (Y = 0.40) and primordial-helium (Y = 0.245) isochrones using the Dartmouth Stellar Evolution Database \citep{2008ApJS..178...89D}.
A detailed description of these helium isochrones is given in \cite{2025ApJ...983...95H}. 
This yields $M_{\rm{He,R}}^{\rm{RGB}}=0.62$~M$_\odot$ and $M_{\rm{He,P}}^{\rm{RGB}}=0.82$~M$_\odot$ for the helium-rich and primordial-helium populations, respectively.
For HB stellar masses, we adopt values from previous studies: $M_{\rm{He,R}}^{\rm{HB}}=0.47$~M$_\odot$ for the helium-rich population \citep{Tailo2015Natur}, and $M_{\rm{He,P}}^{\rm{HB}}=0.62$~M$_\odot$ for the primordial-helium population \citep{2011MNRAS.416L...6M}. 
The resulting mass losses during the red giant phase are therefore $\Delta M_{\rm{He,R}}=0.15$~M$_\odot$ and $\Delta M_{\rm{He,P}}=0.20$~M$_\odot$, implying that the helium-rich population experiences $\sim25$\% less mass loss than the primordial-helium population.
The RGB lifetimes, $\tau_{\rm{He,R}}=520$~Myr and $\tau_{\rm{He,P}}=580$~Myr, are also taken from the Dartmouth database (A. Dotter, priv. comm.). 
From these, the average stellar mass loss rates are $\dot{M}_{\rm{He,R}}=2.81\times10^{-10}$~$M_{\odot} \mathrm{yr}^{-1}$ and $\dot{M}_{\rm{He,P}}=3.44\times10^{-10}$~$M_{\odot} \mathrm{yr}^{-1}$. 
Thus, the helium-rich stars indeed lose mass at a lower rate than primordial-helium stars.

After the above calculations, the total mass-loss rate of \omc{} within the half-light radius can then be calculated using the following equation:
\begin{equation}
    \begin{aligned}
        \dot{M}_{\rm{total}} = N_{\rm{giant}}[f_{\rm{He,R}} \dot{M}_{\rm{He,R}} +(1-f_{\rm{He,R}}) \dot{M}_{\rm{He,P}}]
    \end{aligned}
\label{eqn:masslossrgblife}
\end{equation}
where $N_{\rm{giant}}=13,130$ is the number of giant stars in our sample, and $f_{\rm{He,R}}=44$\% is the fraction of helium-rich stars from \cite{2025ApJ...984..162C}. 
Inputting the above values, we obtain a total mass loss rate of $\dot{M}_{\rm{total}}=4.15\times10^{-6}~M_{\odot}\rm{yr}^{-1}$.
Assuming that hydrogen dominates the ejected gas and that the stars are distributed within a cube of $10'$ per side (i.e., $l=4.87\times10^{19}~\rm{cm}$, matching the oMEGACat footprint), our estimated $\dot{M}_{\rm{total}}$ corresponds to an increase in the column density of $\dot{N}_{\rm{total}}=1.56\times10^{12}~\rm{cm}^{-2} \rm{yr}^{-1}$.

As a caveat to these calculations, our giant star selection is likely to include some asymptotic giant branch (AGB) stars which have very similar luminosities and effective temperatures to first-ascent giant branch stars, but lower mass, and for which the above mass loss calculation is not accurate. Since AGB stars are likely only $\sim 5$--10\% as numerous as RGB stars \citep{Gratton2010}, and in addition only helium-poor AGB stars are likely to show significant post-horizontal branch mass loss, the total mass loss rate is not strongly affected by AGB stars unless their mean mass loss rate is greater than RGB stars by a factor of 10 or more. Unfortunately, it is currently not possible to calculate the mean AGB mass loss rate in the same manner as for RGB stars, as it is challenging to make the mass measurements of AGB stars that would be needed. The fragmentary astroseismic measurements that do exist tend to suggest low masses on the early AGB in metal-poor globular clusters \citep{Howell2025}, which would imply mass loss rates more in line with RGB stars, rather than much larger. This is broadly consistent with mass loss rates estimated for low-metallicity field AGB stars on the early AGB \citep{2009AJ....138.1485D}, which are not too dissimilar from RGB stars of comparable luminosities. We conclude that, while not certain, we do not expect AGB stars to substantially increase our estimate of the total mass loss.

As a comparison to this average mass loss methodology, we also compute the total mass-loss rate using Reimers’ law. 
We adopt a modified version of Reimer’s law calibrated using HB-based measurements of mass loss in globular clusters (Eq.~4 of \citealt{2005ApJ...630L..73S}), which yields a higher value of $6.54\times10^{-6}~M_{\odot}\rm{yr}^{-1}$ compared to our estimate based on the RGB-HB mass difference.

Although the Reimers' law in \citet{2005ApJ...630L..73S} does not include metallicity dependence (potentially introducing systematics of $\lesssim30$\%; \citealt{2015MNRAS.448..502M}), the discrepancy between the two estimates is more likely due to helium enrichment. 
In \omc{}, helium-rich stars lose mass at a lower rate than primordial-helium stars. 
Consequently, this helium-rich population lowers the total stellar mass loss rate. Our central interest is in the total mass loss rate of the stellar population, rather than the star-by-star mass loss rates provided by a modified Reimer’s law. Therefore, the most accurate estimate is that provided by our RGB--HB mass difference calculations, which explicitly account for helium-rich populations and are directly measured for \omc{}. We adopt these for the following discussions.

\subsubsection{Upper Limit on the Atomic Gas Column Density}
\label{subsubsec:discuss-icmgasdensity-atomicgas}

To constrain the intracluster atomic gas column density, we use the uncertainty of the \ewicmdone{} measured from the stacked warm HB stellar spectrum in Fig.~\ref{fig:fitting-icm-stacked} and derive a $3\sigma$ upper limit for the \nad{} absorption: $W_{\mathrm{Na\,\textsc{i}\,D_1,ICM}}$~$=0.0054$~\angstrom{}.
The corresponding sodium atomic gas column density by assuming a covering fraction of 100\% is then estimated using the following equation:
\begin{equation}
    N=2\times 1.13 \times 10^{20} \mathrm{~cm}^{-2} \frac{W}{f \lambda^2},
\label{eqn:icmew2columndensity}
\end{equation}
where $W$ and $\lambda$ are the EW and wavelength of the \nadone{} absorption line in \angstrom{}, respectively, and the oscillator strength $f$ is 0.320 \citep{2015MNRAS.452..511M}. 
The factor of 2 accounts for the fact that the detected atomic gas only represents half of the total column density.
Using this calculation, we obtain an upper limit for the sodium atomic gas column density of $N^{\rm{Na}}_{\rm{atomic}}\lesssim1.09\times10^{11}~\rm{cm^{-2}}$. 
This estimate is comparable to the limit set by \cite{2009MNRAS.399..195V} for \nad{} to be detectable in their spectra.

Assuming the atomic gas is dominated by hydrogen, we estimate the total atomic gas column density by using the following equation:
\begin{equation}
    N_{\rm{atomic}}=\frac{N^{\rm{Na}}_{\rm{atomic}}}{(\mathrm{Na} / \mathrm{H})_{\omega \rm{Cen}}},
\label{eqn:totalcolumndensity}
\end{equation}
where $(\mathrm{Na} / \mathrm{H})_{\omega \rm{Cen}}$ is the number ratio between sodium and hydrogen in \omc{}.
We take the $[\rm{Na}/\rm{Fe}]$ and $[\rm{Fe}/\rm{H}]$ measurements of RGB stars from \cite{2010ApJ...722.1373J} and find an average of $[\rm{Na}/\rm{H}]=-1.53$~dex.
By taking the solar $(\mathrm{Na} / \mathrm{H})_{\odot}=1.7\times10^{-6}$ from \cite{2009ARA&A..47..481A}, we obtain $(\mathrm{Na} / \mathrm{H})_{\omega \rm{Cen}}=5.02\times10^{-8}$.
Then the estimated upper limit of total atomic gas column density is $N_{\rm{atomic}}\lesssim2.17 \times 10^{18}~\rm{cm^{-2}}$. 
The only previous detection of any atomic gas in a GC is by \citet{2006MNRAS.365.1277V}, who detected HI emission from M15 that suggests $\sim$0.3~M$_\odot$ of atomic gas in that cluster.  
Assuming the gas is distributed over a $10'$ cube as above; our upper \nad{} limit translates to an upper limit of 4.3~M$_\odot$; thus we would not be sensitive to the atomic gas detected in M15 by \citet{2006MNRAS.365.1277V}.

\subsubsection{Constraining the Intracluster Medium in \texorpdfstring{\omc}{}}
\label{subsubsec:discuss-icmgasdensity-ionizedandtimescale}


If all the gas in \omc{} produced by stellar mass loss remained atomic, the estimated timescale to accumulate the observed upper limit column density would be approximately 1.4~Myr. 
One common mechanism for gas removal is ram pressure stripping during Galactic disk crossings. 
\cite{2024AA...686A.283P} estimated this crossing timescale for \omc{} to be roughly 40 Myr, which is $\sim$30$\times$ longer than the atomic gas accumulation timescale. 
This discrepancy indicates that atomic gas must be efficiently converted into another phase, or it is driven out of the cluster \citep[e.g.][]{2020MNRAS.491.4602N}.
Previous studies \citep[e.g.,][]{2015MNRAS.446.2226M} have concluded that most ICM in globular clusters should be fully ionized. 
Additionally, the high temperature of red giant winds should be sufficient to ionize hydrogen \citep{2009AJ....138.1485D, 2022ApJ...932...57H}.
Therefore, to constrain the mass loss timescale, it is necessary to estimate the ionized gas content in \omc{}.


We estimate an upper limit on the ionized gas density in \omc{} using published dispersion measure (DM) values of pulsars. 
The dispersion measure is the integral of electrons along the line of sight, and it causes a frequency-dependent delay in the arrival times of pulsar signals.
By combining dispersion measure with pulsar period derivatives ($\dot{P}/P$), we can disentangle the contribution of ionized gas within the cluster from the foreground. 
This method is described in detail by \cite{2001ApJ...557L.105F}, who applied it to 47~Tuc and estimated an intracluster electron density of $n_e=0.067\pm0.015~\rm{cm}^{-3}$.
A follow-up study by \cite{2018MNRAS.481..627A} found a $\sim3\times$ higher density ($n_e=0.23\pm0.05~\rm{cm}^{-3}$) using a larger sample of pulsars and more precise timing measurements. 

\begin{figure}
    \centering
    \includegraphics[width=0.95\linewidth]{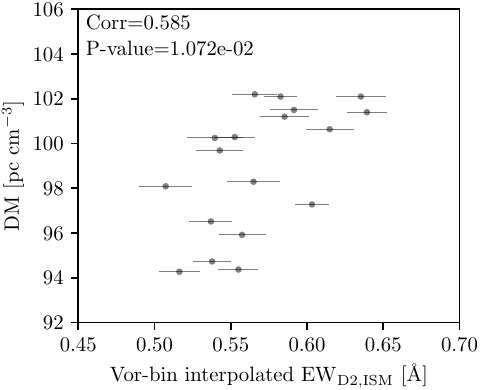}
    \caption{
    Correlation between the dispersion measure and interpolated \ewismdtwo{} from the Voronoi-binned map (right panel of Fig.~\ref{fig:ew-ism-distribution}), using the dispersion measures and coordinates of 18 pulsars \citep{2023MNRAS.520.3847C, 2023MNRAS.521.2616D}.
    This figure shows a strong correlation between the dispersion measure and foreground Na atomic gas, indicating that it is inappropriate to attribute the dispersion measure variation solely to intracluster ionized gas.
    }
    \label{fig:dm-fgnad-relation}
\end{figure}


There are 18 pulsars in \omc{} have published dispersion measures \citep{2023MNRAS.520.3847C, 2023MNRAS.521.2616D} ranging from 94.3 to 102.2~$\rm{pc}~\rm{cm^{-3}}$.
These values are much larger and span a larger range than the dispersion measures of 47~Tuc pulsars ($\sim24.5$~$\rm{pc}~\rm{cm^{-3}}$, \citealt{2018MNRAS.481..627A}).
Given only five of these 18 pulsars have estimates of $\dot{P}/P$, we adopt the simpler approach described in \cite{2001ApJ...557L.105F} to estimate the intracluster ionized gas volumn density of \omc{} by the following equation:
\begin{equation}
    n_e \simeq \frac{\mathrm{std}(\mathrm{DM})}{\mathrm{std}\!\left(\theta_{\perp} \times D\right)},
\label{eqn:ionzedgas}
\end{equation}
where $D = 5.43 \pm 0.05$~kpc is the distance of \omc{} \citep{2021MNRAS.505.5957B}, $\mathrm{std}\mathrm{(DM)}=2.73$~$\rm{pc}~\rm{cm^{-3}}$ is the standard deviation of the dispersion measures of these 18 pulsars, and $\theta_{\perp}$ is the angular distance of each pulsar to the cluster center.
Using this method, we derive an ionized gas density of $n_e = 1.94~\rm{cm}^{-3}$, which is $\sim8\times$ higher than the estimate for 47~Tuc ($0.23\pm0.05~\rm{cm}^{-3}$, \citealt{2018MNRAS.481..627A}).
We note that Eq.~\ref{eqn:ionzedgas} assumes the observed dispersion measure variations arise entirely from ionized gas within the cluster.
However, this may not hold for \omc{} due to its larger range of dispersion measure values.  Furthermore, our findings above that the variations in atomic gas absorption and extinction are almost entirely in the foreground of the cluster suggest this may also be the case for the variations in dispersion measure.  
We find a strong correlation between the dispersion measure values and the foreground \nad{} absorption (\ewismdtwo{}) as shown in Fig.~\ref{fig:dm-fgnad-relation}. 
This correlation is significant at the 99\% level. 
Correcting for this correlation reduces $\mathrm{std}\mathrm{(DM)}$ by $\sim$20\%.
But this likely underestimates the foreground contribution because \nad{} does not trace ionized gas directly.
Therefore, the ionized gas density we derive should be regarded as an upper limit, with a potentially significant fraction of the observed dispersion measure variation arising from foreground ionized gas.
If \omc{} has the same ionized gas density as 47~Tuc, the expected dispersion measure variation would be $\sim$9$\times$ lower than the observed value, $\mathrm{std}\mathrm{(DM)}\sim0.32$~$\rm{pc}~\rm{cm^{-3}}$.

As in Section~\ref{subsubsec:discuss-icmgasdensity-massloss}, we estimate the upper limit of the intracluster ionized gas column density using $N_e = n_e \times l$, which yields $N_e\lesssim9.48 \times 10^{19} \rm{cm}^{-2}$.
Assuming that the intracluster molecular gas content is negligible, we derive an upper limit on the total gas column density (atomic + ionized) of $N_{\rm{total}}\lesssim9.69\times10^{19}~\rm{cm}^{-2}$.
This value is $\sim$50$\times$ higher than the upper limit inferred for atomic gas above.
This upper limit is consistent with the total gas column density range predicted by \cite{2024AA...686A.283P}, which spans from $2.77\times10^{18}$ to $1.30\times10^{20}$~$\rm{cm}^{-2}$.
Their estimate is based on the method of \cite{1975MNRAS.171..467T}, which considers both the lifetimes and the number of HB stars that experienced mass loss during earlier evolutionary stages.
Using the estimates of $N_{\rm{total}}$ and $\dot{N}_{\rm{total}}$, we derive a mass loss timescale upper limit of $\sim60$~Myr. 
This timescale is of the same order of magnitude as the time since \omc{} last crossed the Galactic disk as reported by \cite{2024AA...686A.283P}.
However, given that the ionized gas probed by dispersion measure variation is primarily foreground rather than intracluster, this suggests the gas accumulation timescale could be shorter than the disk crossing time. 
This would imply that the gas is being removed from the cluster through other mechanisms such as stellar winds \citep[e.g.,][]{2011ApJ...728...81M, 2015MNRAS.446.2226M, 2020MNRAS.493.1306C, 2020MNRAS.491.4602N, Zhang2025SciBu}.

\subsection{Comparing Foreground \texorpdfstring{\nad}{} Absorption with Differential Reddening}
\label{subsec:discuss-comparison}

\begin{figure*}
    \centering
    \includegraphics[width=1\linewidth]{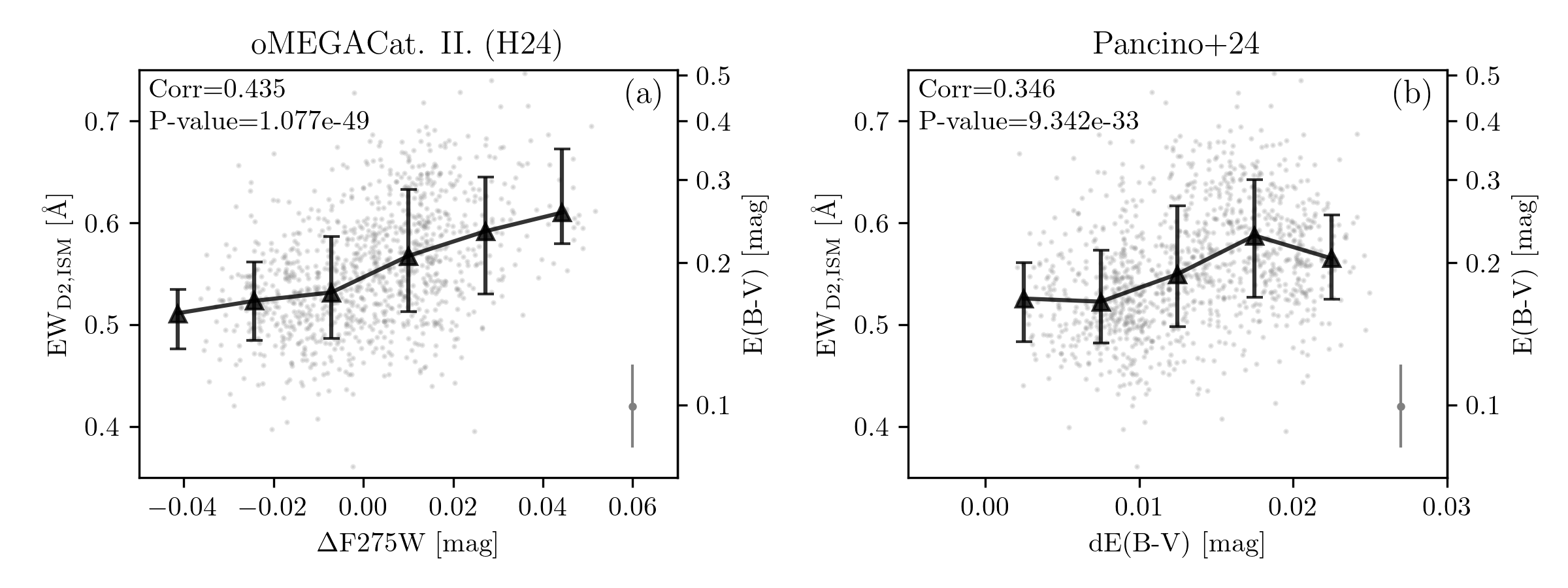}
    \caption{
    Comparison of \ewismdtwo{} with differential reddening measurements from the literature.
    In each panel, gray dots represent individual warm HB stars.
    The left and right y-axes show the \ewismdtwo{} measurements from HB spectra and the inferred \ebv{} using the empirical relation (Eq.~\ref{eqn:ebvewism}), respectively.
    The x-axis indicates differential reddening interpolated from literature maps at the coordinates of the warm HB stars.
    \textbf{Panel (a):} Comparison with photometric correction in the F275W filter from the oMEGACat photometry catalog (\citetalias{2024ApJ...970..192H}). This correction accounts for both differential reddening and CTE degradation, and the F275W band is expected to be dominated by differential reddening.
    \textbf{Panel (b):} Comparison with differential reddening estimates derived from ground-based photometry by \cite{2024AA...686A.283P}.
    Black triangles indicate the median of each x-axis bin, with the ranges obtained from $16^{th}$ and $84^{th}$ percentiles.
    The Pearson correlation coefficient and p-value are listed in each panel.
    A median uncertainty of the \ewismdtwo{} measurements is shown in the lower right in gray.
    The mild correlation supports the conclusion that sodium atomic gas is a reliable tracer of dust, and that the observed substructures in differential reddening are primarily due to foreground ISM.
    }
    \label{fig:fitting-ism-lit}
\end{figure*}

Given the strong correlation between sodium atomic gas and dust \citep[e.g.,][]{2009MNRAS.399..195V, 2012MNRAS.426.1465P, 2015MNRAS.452..511M}, the substructures seen in our foreground \nad{} EW distribution are expected to be consistent with those observed in differential reddening distributions from previous studies \citep[e.g.,][]{2017ApJ...842....7B,2024AA...686A.283P,2024ApJ...970..192H}.  
In this section, we quantitatively compare our \ewismdtwo{} distribution with differential reddening from the literature to verify this consistency.

The left panel of Fig.~\ref{fig:fitting-ism-lit} shows our \ewismdtwo{} measurements of warm HB spectra against the photometric corrections in the F275W filter from the oMEGACat photometric catalog \citepalias{2024ApJ...970..192H}.
\citetalias{2024ApJ...970..192H} computed photometric corrections for all HST filters to account for both differential reddening and systematic zero-point variations caused by the decreasing charge transfer efficiency (CTE) of HST’s ageing detectors.
Because the F275W filter is at short wavelengths and highly sensitive to dust extinction, its photometric correction is expected to be dominated by differential reddening.
Therefore, we use the photometric correction in $\Delta\mathrm{F275W}$ as a tracer of differential reddening in this comparison.
The right panel compares \ewismdtwo{} results with differential reddening values from ground-based photometry by \cite{2024AA...686A.283P}. 
In both panels, the differential reddening was obtained by measuring the distance of reference stars to fiducial lines in the CMD. 
The x-axes show differential reddening values interpolated at the positions of the warm HB stars in our sample. 
The mild correlation and small p-values in both panels confirm the robust relation between sodium atomic gas and dust.

Since our \ewismdtwo{} measurements reflect only the contribution from the foreground gas, this confirms that the observed differential reddening substructures in \omc{} from previous studies are due to foreground rather than intracluster dust. 
Together with our findings in Section~\ref{sec:results-icm}, which show no clear substructures in the \ewicmdone{} distribution, and those in Section~\ref{subsubsec:discuss-icmgasdensity-ionizedandtimescale}, where pulsar dispersion measures correlate with foreground \nad{}, our results suggest that the observed variations in both dust and ionized gas toward \omc{} are primarily attributable to foreground material.
This consideration should also apply to other globular clusters at low Galactic latitudes with significant foreground extinction.



\subsection{Mismatch in Differential Reddening Estimates}
\label{subsec:discuss-dr-mismatch}

Despite the correlation between \ewismdtwo{} and the inferred photometric differential reddening, the data points in Fig.~\ref{fig:fitting-ism-lit} are very scattered. 
This scatter could come from either measurement errors (on the EW measurement or the photometric reddening estimate) or an intrinsic scatter in the relationship between \nad{} and reddening.  
To investigate this further, we compare the differential reddening estimated from our \nad{} measurements vs. photometric measurements.  
This indicates that the observed scatter is physical and not dominated by measurement errors, reflecting local variations in the relation between atomic gas and dust extinction.

In Section~\ref{subsec:results-ism-dr}, we derive a differential reddening estimate from \ebv{} values inferred using the empirical \nad{}-reddening relation of Eq.~\ref{eqn:ebvewism}. 
This measurement accounts for \ewismdtwo{} measurement uncertainties. 
However, the resulting differential reddening still includes intrinsic scatter in the \nad{}-reddening relation. 
As shown in Table~\ref{tab:reddening_dr}, the resulting differential reddening values both for warm HB stars and all other stars samples (first two rows) are higher than those reported in the literature, and also higher than the value derived from color displacement of warm HB stars relative to the fiducial sequence (third row). 
Therefore, these results provide strong evidence of significant scatter in the \nad{}-reddening relation.

Intrinsic scatter in the \nad{}-reddening relation has been suggested in previous studies. 
\cite{2009MNRAS.399..195V} compared the EW of foreground \nad{} absorption using HB stars of \omc{} with reddening values from DIRBE/IRAS data \citep{1998ApJ...500..525S} and reported similar scatter. 
Furthermore, significant scatter is also seen when \cite{2012MNRAS.426.1465P} derived the empirical relation of Eq.~\ref{eqn:ebvewism} using QSO spectra (see their Fig.~5). 
\cite{2009MNRAS.399..195V} attributed this scatter primarily to variations in ionization equilibrium. 
Our observations covering such a narrow field of view strongly suggest that these variations occur over small scales in the Milky Way's ISM.

To evaluate the effect of \nad{}-inferred reddening corrections on the CMD, \cite{2009MNRAS.399..195V} applied these corrections to their horizontal branch stars and found that the HB sequence became more scattered (see their Fig.~12).
We perform a similar test on our warm HB stars and obtain consistent results: the HB sequence becomes wider after applying the \nad{}-based corrections.
In contrast, the sequence becomes narrower when corrected using the F275W-based photometric maps from \citetalias{2024ApJ...970..192H}.
This supports the interpretation that our \nad{}-based differential reddening estimates are overestimated due to the scatter in the \nad{}-reddening relation.
Since the empirical relation from \cite{2012MNRAS.426.1465P} is derived from QSO spectra along many lines of sight, the observed scatter in this relation indicates that such small-scale fluctuations in ionization equilibrium occur throughout the Galaxy.

\begin{figure}
    \centering
    \includegraphics[width=1\linewidth]{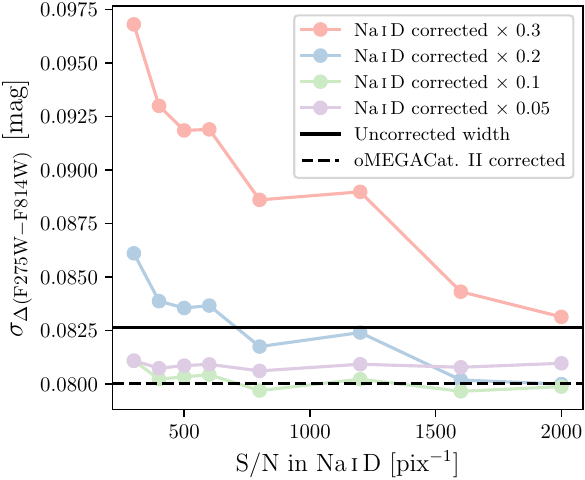}
    \caption{
    Width of horizontal branch (y-axis) from \nad{}-corrected photometry in different relative reddening rescaling factors and Voronoi binning S/Ns. 
    The solid line represents the HB width from uncorrected photometry in \citetalias{2024ApJ...970..192H}.
    The dashed line shows the HB width after applying the corrections from \citetalias{2024ApJ...970..192H}.
    Colored lines indicate HB widths after correcting the photometry with our \nad{}-inferred relative reddening scaled by various factors.
    This figure demonstrates that applying a rescaling factor of 0.1 to our \nad{}-inferred reddening yields an HB width comparable to that obtained from \citetalias{2024ApJ...970..192H}’s correction.
    }
    \label{fig:hb-cmd-thickness}
\end{figure}

We also test whether applying a scaling factor to the variation in \nad{}-inferred reddening improves CMD corrections. 
The results are shown in Fig.~\ref{fig:hb-cmd-thickness}, where we plot the HB sequence width (y-axis) as a function of Voronoi binning S/N. 
The individual warm HB star \ewismdtwo{} estimates have significant individual measurement errors, so we use the Voronoi-binned maps instead. 
We interpolate reddening values at the positions of warm HB stars using Voronoi-binned \ewismdtwo{} maps at various S/N requirements of Voronoi binning. 
We vary the S/N to see the impact of the \ewismdtwo{} measurement errors; at higher S/N these become very small. 
The relative reddening (defined as the deviation from the median) is then scaled by factors of 0.3, 0.2, 0.1, and 0.05 and applied to correct the photometry. 
The HB sequence width is measured as the standard deviation of each star’s color offset from the fiducial sequence. 
For comparison, we include the HB widths from the uncorrected (solid black line) and F275W-corrected (dashed black line) photometry in \citetalias{2024ApJ...970..192H}. 
From this figure, we find that applying a scaling factor to the \nad{}-inferred relative reddening can reduce the HB width.
In particular, by using a factor of 0.1, we can obtain a comparable HB width to that obtained from the oMEGACat photometric correction.
This factor quantitatively reflects how much our differential reddening is overestimated.

Other factors may also contribute to the observed mismatch in differential reddening. 
For example, the \nad{}-reddening relation from \cite{2012MNRAS.426.1465P} is poorly constrained in the EW range relevant to our measurements. 
As shown in their Fig.~9, most of their data points have EW values below 0.4~\AA{}, with only one data point in the 0.4-0.7~\AA{} range of our \ewismdtwo{} measurements. 
This sparse sampling may fail to capture a steepening trend in this regime, potentially leading to an overestimated range of inferred reddening. 
Additionally, photometric uncertainties in previous studies, such as charge transfer efficiency losses or crowding in ground-based imaging, can influence the photometric differential reddening measurements \citep[see Section 3.2 of][]{2012A&A...540A..16M}. 
The choice of smoothing strategy in photometric methods may further affect these estimates. 
Nevertheless, given the significant scatter observed in the \nad{}-reddening relation (Fig.~\ref{fig:fitting-ism-lit}), we consider it the primary cause of the differential reddening mismatch.

In conclusion, the mismatch between our \nad{}-inferred differential reddening estimates and literature values can be attributed to physical scatter in the \nad{}-reddening relation, likely caused by ionization equilibrium variations in the ISM. 
By rescaling the \nad{}-inferred relative reddening, we can correct the CMD photometry and recover HB sequence widths comparable to those from photometric differential reddening estimates.

\subsection{Connection between Foreground ISM and \texorpdfstring{\omc}{}}
\label{subsec:discuss-icm-omc-relation}

\begin{figure}
    \centering
    \includegraphics[width=1\linewidth]{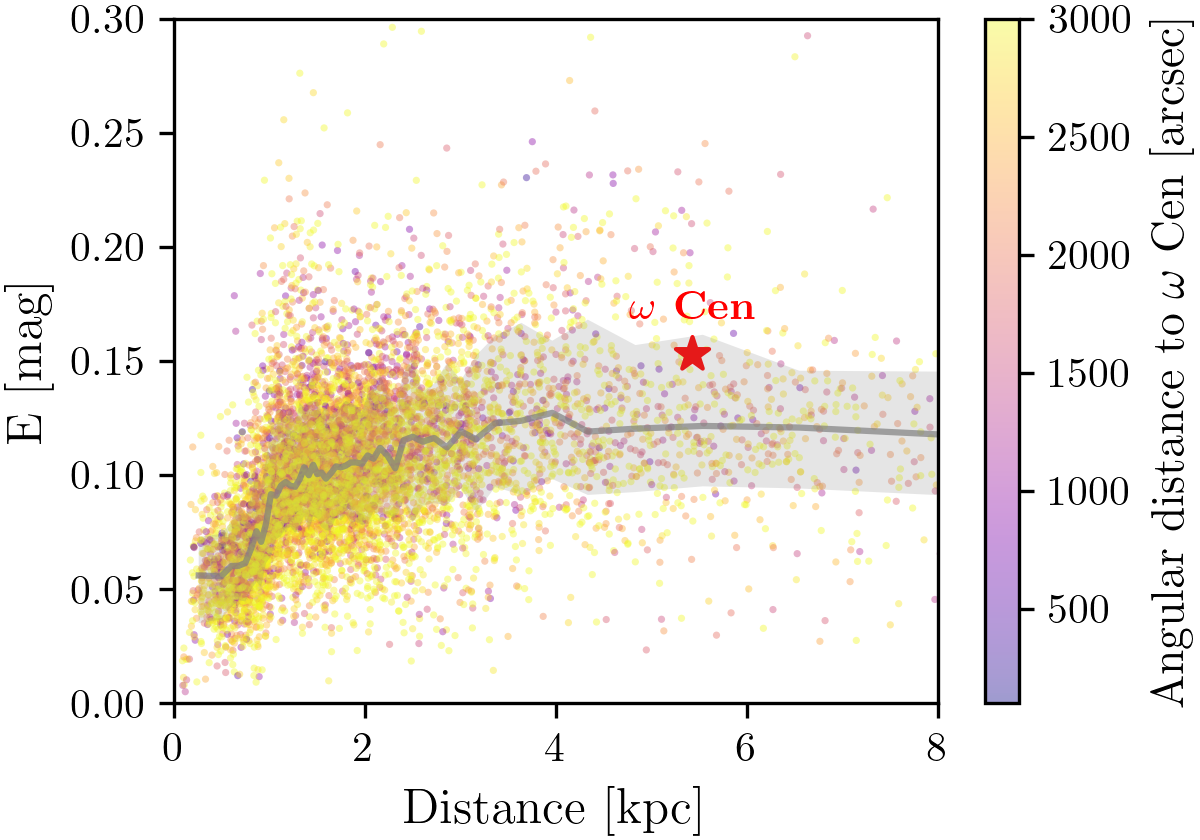}
    \caption{
    Reddening ($E$) as a function of distance for stars within 1 degree of the \omc{} from Gaia DR3.
    Stars are color-coded by their angular distance to the center of \omc{}. 
    The gray line and shaded region represent the median and ($16^{th}$, $84^{th}$) percentiles.
    The reddening and distance values are derived from Gaia BP/RP spectra using a machine learning method \citep{2023MNRAS.524.1855Z}. 
    The distance and measured reddening of \omc{} are denoted by a red star marker.
    This figure indicates that the ISM is primarily located at distances between 0-2 kpc. Therefore, the foreground ISM is unlikely to interact with \omc{}.
    }
    \label{fig:ism_gaiaxp}
\end{figure}

One interesting finding from the \ewismdtwo{} spatial distribution in Fig.~\ref{fig:ew-ism-distribution} is that the inner region within a radius of 100~arcsec demonstrates lower EWs than the outer regions. 
This is also observed in the differential reddening distributions from previous studies on \omc{} (e.g., \citetalias{2024ApJ...970..192H} and \citealt{2024AA...686A.283P}). 
As discussed in Section~\ref{subsec:discuss-comparison}, we have ruled out the ICM origin of this feature. 
Therefore, in this subsection, we investigate whether this feature is due to the interaction between \omc{} and the foreground ISM by exploring the reddening distribution along the line of sight.

To investigate the distance of the ISM toward \omc{}, we present in Fig.~\ref{fig:ism_gaiaxp} the extinction distribution as a function of distance to the Sun using stars within 1 degree of the \omc{} center from Gaia DR3 \citep{2023A&A...674A...1G}.
Stars in this figure are non-cluster members with proper motions larger than $3$~\maspyr{} relative to \omc{}, and they are color-coded by the angular separation from the \omc{} center. 
The gray line and shaded region represent the median and $1\sigma$ trends.
The reddening and distance values are derived from Gaia BP/RP spectra combined with 2MASS \citep{2006AJ....131.1163S} and WISE \citep{2019ApJS..240...30S} photometry using a machine learning approach developed by \cite{2023MNRAS.524.1855Z}.
This model was trained using stellar atmosphere parameters from LAMOST, Gaia DR3 parallaxes \citep{2023A&A...674A...1G}, and reddenings from the 3D dust map Bayestar19 \citep{2019ApJ...887...93G}. \cite{2023MNRAS.524.1855Z} also noted that their estimated reddening value $E$ is comparable to \ebv{}.
We apply a quality cut by excluding stars with \texttt{xp\_quality\_flags}~$>0$, which removes stars with poor fitting quality (see details in \citealt{2023MNRAS.524.1855Z}). 
This figure shows that the foreground medium is primarily located at distances between $0-2$~kpc, beyond which the extinction distribution becomes flat.
Based on \omc{}’s low Galactic latitude, this implies that the foreground medium observed in Fig.~\ref{fig:ew-ism-distribution} is mainly in the Galactic plane, and thus is not located close to \omc{}.
Therefore, the smaller EW of \nadtwo{} line in the central region of \omc{} is likely just an observational coincidence.

Our conclusion is consistent with previous findings by \cite{1992MNRAS.254..221B, 1993ApJ...417..572W, 1994MNRAS.267..660W}, which identified strong foreground ISM absorption as part of a vertical extension of the Carina-Sagittarius spiral arm.
Our MUSE spectra measurements report the mean \vlos{} of foreground sodium gas is $-13\pm5$~\kms{}, which is also consistent with their results of $-15$~\kms{}. 
This also agrees with the \cite{2009MNRAS.399..195V} measurement of $-14$~\kms{}, where they concluded that the \nad{} lines originate predominantly from a puffed-up spiral arm at 1-2 kpc.

\section{Summary} 
\label{sec:summary}

In this study, we investigate atomic gas associated with the ISM and ICM of \omc{} using \nad{} absorptions in MUSE spectra.
By measuring the EWs of \nad{} lines at different line-of-sight velocities, we separate the contributions from foreground and intracluster gas.

For the foreground atomic gas (\vlos{} at $-13\pm5$~\kms{}), we identify substructures in its spatial distribution using spectra from both warm HB stars and Voronoi-binned non-warm-HB (mostly MS to RGB) stars. 
The outer regions exhibit larger \nad{} EWs compared to the central 100~arcsecs.
By analyzing the reddening distribution along the line of sight using Gaia DR3 \citep{2023A&A...674A...1G}, we find that this trend may be coincidental, as most foreground dust is located within $0-2$~kpc of the Sun, well in front of \omc{}.

Using the \nad{}-reddening empirical relation from \cite{2012MNRAS.426.1465P}, we transfer the EW of foreground \nad{} absorptions to reddening and estimate an average \ebv{} of $0.153 \pm 0.003$~mag within the half-light radius of \omc{}.
We also find a mild correlation ($\rm{Corr=0.435}$) between our foreground \nad{} EW and differential reddening derived from photometric methods \citep[e.g.,][]{2017ApJ...842....7B, 2024ApJ...970..192H, 2024AA...686A.283P}, confirming that \nad{} traces dust.
However, the correlation exhibits significant scatter beyond the EW uncertainties.
Applying an uncertainty deconvolution method to the \nad{}-inferred reddening yields a differential reddening of $dE(B-V) = 0.026 \pm 0.003$~mag, which is much larger than values reported in previous photometric studies.
We attribute this mismatch to the scatter in the \nad{}-reddening relation, likely caused by local fluctuations in gas ionization equilibrium and extinction.
Despite this, we find that scaling the relative \nad{}-inferred reddening by a factor of 0.1 allows us to correct the photometry of warm HB stars in the CMD, yielding a sequence thickness comparable to that obtained using photometric differential reddening estimates.

For the intracluster atomic gas, we detect no significant \nad{} absorption at \omc’s systemic velocity (232.7~\kms{}).
Individual measurements of warm HB stars show no spatial substructures or radial gradients in the \nad{} distribution, confirming that all observed substructures in the differential reddening map originate from the foreground.  This result calls into question the potential detection of ICM in Milky Way globular clusters by \citet{2024AA...686A.283P} using differential reddening, especially for clusters at lower Galactic latitude like \omc{}.  These ICM detections could be tested with future \nad{} based studies.

From the stacked warm HB spectrum, we measure the EW of \nadone{} from ICM as $0.0007 \pm 0.0018$~\angstrom{}; this limit is much lower than the previous \nad{} ICM constraints placed by \citet{2009MNRAS.399..195V}.
This yields an upper limit on the intracluster atomic gas column density of $N_{\rm{atomic}} \lesssim 2.17 \times 10^{18}~\rm{cm^{-2}}$.
We also find that the dispersion measure variations among pulsars in \omc{} are correlated with the foreground \nad{} absorption.  This suggests that the inferred intracluster ionized gas density from these pulsars of 
$N_e \lesssim 9.48 \times 10^{19} \rm{cm}^{-2}$, is an upper limit. This provides direct evidence that the gas is being removed from the cluster by internal processes.

In the future, the method in this work can be applied to other globular clusters observed with MUSE \citep{2018MNRAS.473.5591K} to disentangle ISM and ICM contributions to \nad{} absorption, especially in clusters where ICM signatures have been previously reported.
This will enable us to explore the distribution of atomic gas in globular clusters and better understand their formation mechanisms. 
To fully understand different phases of ICM gas, observations with instruments such as ALMA will also be useful to constrain the molecular gas content in \omc{}. 
Additional pulsar observations with more precise timing measurements, or detections of $H_\alpha$ emission, are also needed to better constrain the ionized gas density.
Furthermore, the scatter in the correlation between \nad{} absorption and reddening requires further investigation, either by refining the empirical \nad{}-reddening relation or by exploring fluctuations in ionization equilibrium, especially within the EW range relevant to \omc{}. 
Mapping the spatial distribution of ionization fluctuations would in turn help improve the \nad{}-reddening relation and improve the accuracy of reddening measurements in globular clusters.


\begin{acknowledgments}
We thank 
M. Bergemann, 
J. Dalcanton, 
A. Dotter, 
S. Dreizler, 
Y. Wang for useful discussions.
ZW, ACS, and CC acknowledge support from a \textit{Hubble Space Telescope} grant GO-16777.
JS acknowledges support from NASA grant 80NSSC21K0628.
This work is done using \texttt{Yoga} (\url{https://yoga-server.github.io/}), a privately built Linux server for astronomical computing.
\end{acknowledgments}

%

\vspace{5mm}
\facilities{HST (STScI), VLT/MUSE (ESO)}


\software{Astropy \citep{2013A&A...558A..33A, 2018AJ....156..123A}, Numpy \citep{2020Natur.585..357H}, Scipy \citep{2020NatMe..17..261V}, Emcee \citep{2013PASP..125..306F}, Matplotlib \citep{2007CSE.....9...90H}}

\bibliography{paper}{}
\bibliographystyle{aasjournal}



\end{CJK}
\end{document}

%% file: ewd2ism_table.tex
\begin{deluxetable*}{lll}
\tabletypesize{\footnotesize}
\tablecaption{Equivalent widths of \nadtwo{} (ISM) and \nadone{} (ICM) absorption measured from MUSE spectra of warm HB stars.}
\tablehead{
\colhead{Column} & \colhead{Description} & \colhead{Unit}
}
\startdata
RA          & Right Ascension from \citetalias{2010ApJ...710.1032A} & deg \\
DEC         & Declination from \citetalias{2010ApJ...710.1032A} & deg \\
IDAvdM10    & Identifier from \citetalias{2010ApJ...710.1032A} &  \\
Teff        & Effective temperature obtained from BASTI-ZAHB tracks (see Section~\ref{subsubsec:data-target-selection-hb}) & K \\
SNRNaD      & Re-estimated signal-to-noise ratio  of the MUSE spectrum in the \nad{} region (\snrnad{}) & \perpixel{} \\
RVel        & Line-of-sight velocity of the star & km s$^{-1}$ \\
EWD2ISM$^a$    & Equivalent width of \nadtwo{} (ISM) absorption & \AA{} \\
e\_EWD2ISM$^a$ & Uncertainty of \nadtwo{} (ISM) absorption & \AA{} \\
RVelD2ISM$^a$   & Line-of-sight velocity of \nadtwo{} (ISM) absorption & km s$^{-1}$ \\
EWD1ICM$^b$     & Equivalent width of \nadone{} (ICM) absorption & \AA{} \\
e\_EWD1ICM$^b$ & Uncertainty of \nadone{} (ICM) absorption & \AA{} \\
RVelD1ICM$^b$   & Line-of-sight velocity of \nadone{} (ICM) absorption & km s$^{-1}$ \\
chi2        & $\chi^2$ of the spectral fitting &  \\
\enddata
\tablecomments{(a) Only available for stars with \snrnad{}~$>40$~\perpixel{} and $\chi^2<15$, see details in Section~\ref{subsec:results-ism-2d}; (b) Only available for stars with \teff{}~$>10000$~K and \snrnad{}~$>40$~\perpixel{}, see details in Section~\ref{sec:results-icm}. \\ This table lists the columns of the machine-readable table. The full table is available in the online article.}
\label{tab:ewd2ism_table}
\end{deluxetable*}

%% file: reddening_table.tex
\begin{table*}
    \caption{Summary of differential reddening with the methods, data, and FoV from this work and literature.}
    \label{tab:reddening_dr}
    \begin{tabular}{lcccccc}
    \hline\hline
    Sample & Method & Data & FoV & $dE(B-V)$ & Reference \\\relax
    & & & $[\rm{deg}^2]$ & [mag] \\\hline
    Warm HB stars & \nad{} & MUSE & $0.25\times0.16$ & $0.026\pm0.003$ & this work \\
    All other stars & \nad{} & MUSE & $0.25\times0.16$ & $0.035\pm0.001$$^{a}$ & this work \\
    Warm HB stars & photometric$^{b}$ & HST & $0.25\times0.16$ & $0.018\pm0.003$ & this work \\
    HB stars & $[c]$ index$^{c}$ & NTT & $0.23\times0.23$  & 0.030 & (1) & \\
    Main-sequence & photometric & HST & $0.07\times0.07$ & 0.005 & (2) \\
    MS-to-RGB & photometric$^{d}$ & HST & $0.18\times0.18$ & 0.007 & (3) \\
    Main-sequence & photometric & $\begin{array}{c}\text{ground-based} \\ {\text{photometry}^{e}}\end{array}$  & $0.18\times0.18$ & 0.013 & (4) \\
    \hline\hline
    \end{tabular}
    \tablecomments{\footnotesize (a) Results from Voronoi binned residual spectra; (b) Use photometry in F275W and F814W; (c) A reddening-free temperature index calculated by photometric colors; (d) Use the photometric correction map in F275W; (e) Using all the archival photometric observations by ground-based telescopes on \omc{}.}
    \tablerefs{\footnotesize (1) \cite{2005ApJ...634L..69C}; (2) \cite{2017ApJ...842....7B}; (3) \cite{2024ApJ...970..192H}; (4) \cite{2024AA...686A.283P}}
\end{table*}


